\PassOptionsToPackage{table}{xcolor}

\documentclass[twocolumn]{aastex631}

\usepackage{makecell}
\usepackage{amsmath}

\shorttitle{PyVesta: A new Versatile Echelle Spectra exTraction and Analysis pipeline}
\shortauthors{Stock et al.}

\begin{document}
\title{PyVesta: A new Versatile Echelle Spectra exTraction and Analysis pipeline}
\author[0000-0003-2789-7367]{Lukas Stock*}
\affiliation{Marburg University, Marburg, Hessen, Germany}
\correspondingauthor{Lukas Stock}
\email{lukas.stock@physik.uni-marburg.de}

\author[0000-0001-5860-4139]{Andreas Schrimpf}
\affiliation{Marburg University, Marburg, Hessen, Germany}

\begin{abstract}
We present PyVesta, an automatic data reduction software for fiber fed echelle spectrographs. PyVesta is designed to be used with different instruments, both high resolution spectrographs such as HARPS or FEROS as well as inexpensive off-the-shelf spectrographs like the Shelyak eShel II. Special additions were implemented to improve the stability and resolution of inexpensive instruments especially. PyVesta provides tools for all extraction steps from raw CCD images to continuum normalized spectra, e.g., CCD image corrections, spectra extraction, wavelength calibration, and continuum normalization. Our software is based on the software package CERES and was ported to Python 3, revised, and supplemented. 
\end{abstract}


\section{Introduction} \label{sec:intro}
Echelle spectrographs offer high resolution spectra combined with large spectral coverage. That makes them ideal for many use cases, such as spectral classification of stars or the determination of companions via the radial velocity method. Most dedicated exoplanet searching instruments such as HARPS or CARMENES are echelle spectrographs \citep{Mayor2003, Quirrenbach2014}. 

The main disadvantage of echelle spectrographs in comparison to single order spectrographs is the complexity of data reduction for those instruments. Echelle spectra feature multiple diffraction orders, which are shifted relative to each other on the detector chip. These orders are also often curved, which requires precise tracking. Many spectrographs feature multiple fibers, e.g. for calibration spectra, and image slicers, which complicates the extraction even more. In addition, less expensive spectrographs in particular exhibit optical errors in their images, which must be taken into account during extraction. 

Many data processing routines were developed especially for high resolution spectrographs. For example, \cite{Sosnowska2015} developed a software package to extract the data from ESPRESSO and also from HARPS and HARPS-N. In the past, ESO`s MIDAS package \citep{Banse1983} was used to extract echelle spectra from most of the ESO spectrographs such as HARPS, UVES and XSHOOTER \citep{Ballester1992}. IRAF/PyRAF \citep{Tody1986, SSBS2012} is an old but powerful software package for analyzing data from various instruments, but it is now practically out of use as well. PyReduce \citep{Piskunov2021} is a port of the REDUCE software \citep{Piskunov2002} to Python, while PypeIt \citep{pypeit:joss_pub, pypeit:zenodo} provides a modern and modular data reduction pipeline. PyReduce and PypeIt are both well established, (Py)Reduce is used with fiber-fed and slit echelle spectrographs (e.g. \cite{Kochukhov2015, Janson2025}), PypeIt with a wide range of slit and multi-object instruments (e.g. \cite{Tinyanont2024, Carvalho2024}). Many instruments also come with extraction software that has been specifically developed for that particular instrument, e.g., \cite{Bochanski2009, Chakraborty2014, Georgiev2025}. In addition to these large instruments, there is also software for smaller instruments, e.g., the Audela\footnote{\url{https://audela.org/dokuwiki/doku.php/en/start}} package. 

The various software packages differ in many details such as spectrum extraction or wavelength calibration. Using different packages in projects that require data from multiple instruments can lead to systematic errors, especially for high resolution radial velocity analysis. For this reason, MIDAS supports several ESO instruments, but cannot be used for other instruments. These packages are usually written in very specific programming languages and/or development environments, which makes it difficult to adapt them to other instruments. An approach to analyzing data from a number of instruments using only one pipeline was presented by \cite{Brahm2017} with the CERES software package. This package provides the possibility to work with different instruments , e.g. CORALIE, FEROS and HARPS. However, this software was originally written in Python 2, which reached its end of life in 2020. Also new extraction and wavelength calibration routines were published in the meantime by PyReduce \citep{Piskunov2021} and xwavecal \citep{Brandt2020, MirekBrandt2021}, respectively. Therefore, we developed PyVesta, the spiritual successor to CERES. This new software package is written in Python 3 completely, so that adjustments can be done much easier compared to some older or less common programming languages. In addition, PyVesta was specifically designed to extract data from off-the-shelf spectrographs. These spectrographs are generally much less well characterized than high-resolution instruments, which means they benefit significantly from PyVesta's more flexible data reduction. A uniform data reduction using PyVesta also allows data from off-the-shelf and high-resolution spectrographs to be combined more effectively.

\section{Software structure} \label{sec:structure}
\begin{figure*}[tb!]
\centering
\includegraphics[width = \textwidth]{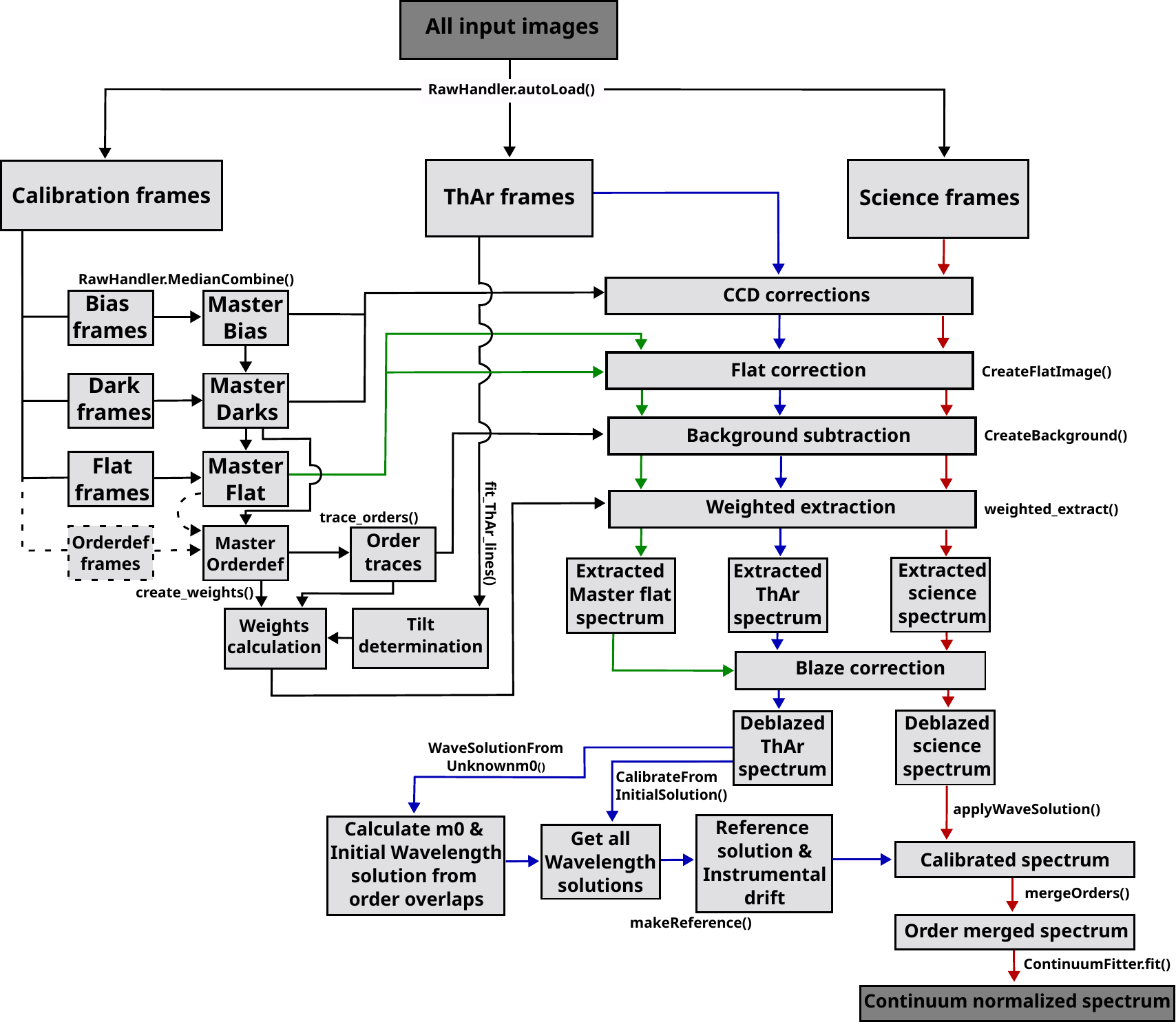}
\caption{Overview of the individual steps in PyVesta. Particularly important functions are mentioned. Orderdef frames are not strictly necessary. If none are available, the master flat will be used instead. A rough overview of how long the key steps take can be found in Table \ref{tab:Extraction_times}. \label{fig:software_structure}}
\end{figure*}
\subsection{General considerations} \label{subsec:structure_general}
The main goal for the development of PyVesta  is to make it compatible with the data of almost all echelle spectrographs. There are only a few conditions that must be satisfied for PyVesta, namely the different orders have to be clearly distinguishable, the wavelength calibration is done using ThAr lamps and that the diffraction orders lie approximately parallel to one another. A general overview of PyVesta's architecture is shown in Fig. \ref{fig:software_structure}. Table \ref{tab:Extraction_times} provides an overview of the approximate time required for each step.

While PyVesta should be compatible to all instruments which are compatible with CERES, we tested our software with HARPS and FEROS in particular. Special attention was paid to our own spectrograph, a Shelyak eShel II. This instrument was described in detail in \cite{Stock2025}. Additional instruments can be easily added. Instructions are available on the PyVesta website\footnote{\url{https://github.com/lukstock/pyvesta/tree/main}}.

\begin{table*}[t!]
\centering
{%
\begin{tabular}{l|c|c|c||l|c|c|c}
\textbf{Step} & \textbf{eShel II} & \textbf{HARPS} & \textbf{FEROS} & \textbf{Step} & \textbf{eShel II} & \textbf{HARPS} & \textbf{FEROS} \\
\hline
\hline
\rowcolor{gray!45}[\tabcolsep][\tabcolsep]
\textbf{Master frames} & $10$ s & $40$ s & $10$ s & \textbf{Overlap wavelength solution } & $150$ s & $600$ s & $180$ s \\
\hline
\rowcolor{gray!5}[\tabcolsep][\tabcolsep]
\textbf{Order tracing} & $20$ s & $80$ s & $150$ s &  \vtop{\hbox{\strut \textbf{Wavelength solution from}} \hbox{\strut \textbf{initial solution}}} & $90$ s & $50$ s & $180$ s \\
\hline
\rowcolor{gray!45}[\tabcolsep][\tabcolsep]
\textbf{Flat image} & $50$ s & $135$ s & $120$ s & \textbf{Background calculation} & $2$ s & $5$ s & $5$ s\\
\hline
\rowcolor{gray!5}[\tabcolsep][\tabcolsep]
\textbf{Line tilt} & $5$ s & - & - &   \vtop{\hbox{\strut \textbf{Optimal extraction}} \hbox{\strut \textbf{of one frame}}} & $40$ s & $140$ s & $60$ s \\
\hline
\rowcolor{gray!45}[\tabcolsep][\tabcolsep]
\textbf{Weights} & $60$ s & $220$ s & $160$ s &  \vtop{\hbox{\strut \textbf{Complete reduction}} \hbox{\strut \textbf{of one frame}}} & $50$ s & $160$ s & $65$ s
\end{tabular}%
}
\caption{Overview of the time required for each step in PyVesta. The values are rounded and are based on a 2023 laptop processor with 8 physical cores and 16 parallel threads. PyVesta was generally allowed to use 12 threads and 2 threads during extraction. The times for “wavelength solution from initial solution,” “Optimal extraction”, and “complete reduction” are given per file. All other steps are required only once per run. Since there is no measurable line tilt in HARPS and FEROS, this calculation is not performed in those cases. For HARPS, it should also be noted that only one of the two components is extracted per run, since PyVesta treats HARPS blue and HARPS red as two separate instruments. \label{tab:Extraction_times}}
\end{table*}

\subsection{Adaptations to new instruments} \label{subsec:structure_adaptions}
PyVesta was designed to be as modular as possible so that it can be easily adapted to different instruments. All instrument-specific settings --- such as the presence of an image slicer or the polynomial degree for various fits during extraction --- are defined in a specific instrument class in the \textit{instruments.py} file. Within this class, small functions need to be adapted that allow PyVesta to read relevant information such as exposure time or MJD from the header. 

Furthermore, all parameters that are not specific to the instrument but rather to the current extraction session, such as which images to save or how many processor cores PyVesta is allowed to use, are stored in a reduction parameters class. These classes are defined in the \textit{config.py} file. Both files already contain configurations for HARPS, FEROS, and the Shelyak eShel II, which can be used as templates.

\section{Image processing} \label{sec:image_processing}
\subsection{Correction frames} \label{subsec:structure_correctionframes}
In addition to stellar spectra, which are referred as science frames, different kinds of calibration data are needed. Bias and dark calibration frames are necessary for CCD corrections, although dark frames can be disregarded for cryogenically cooled CCD detectors. Flat frames are also required for calibration. However, they are used in a different way than in photometry as described in Sec. \ref{subsec:science_blaze}. Two further special calibration frames are required for spectroscopy, namely order definition (orderdef for short) and wavelength calibration frames. Orderdef frames provide a spectrum as flat as possible in order to accurately determine the position of the diffraction orders on the detector. Flat images can also be used for this task, but a special orderdef image is often taken, usually with an additional light source to achieve better illumination across all wavelengths. Thorium-Argon (ThAr) lamps are used for wavelength calibration. These lamps have an emission spectrum with many lines across the visible wavelength range. We will therefore also refer to the wavelength calibration images as ThAr frames. 

The different types of frames are specified in the FITS headers. PyVesta creates a separate subdirectory for each individual frame type in the output directory so that the different types can be easily distinguished.

\subsection{CCD corrections} \label{subsec:structure_CCDcorrections}
\begin{figure*}[t!]
\centering
\includegraphics[width = \textwidth]{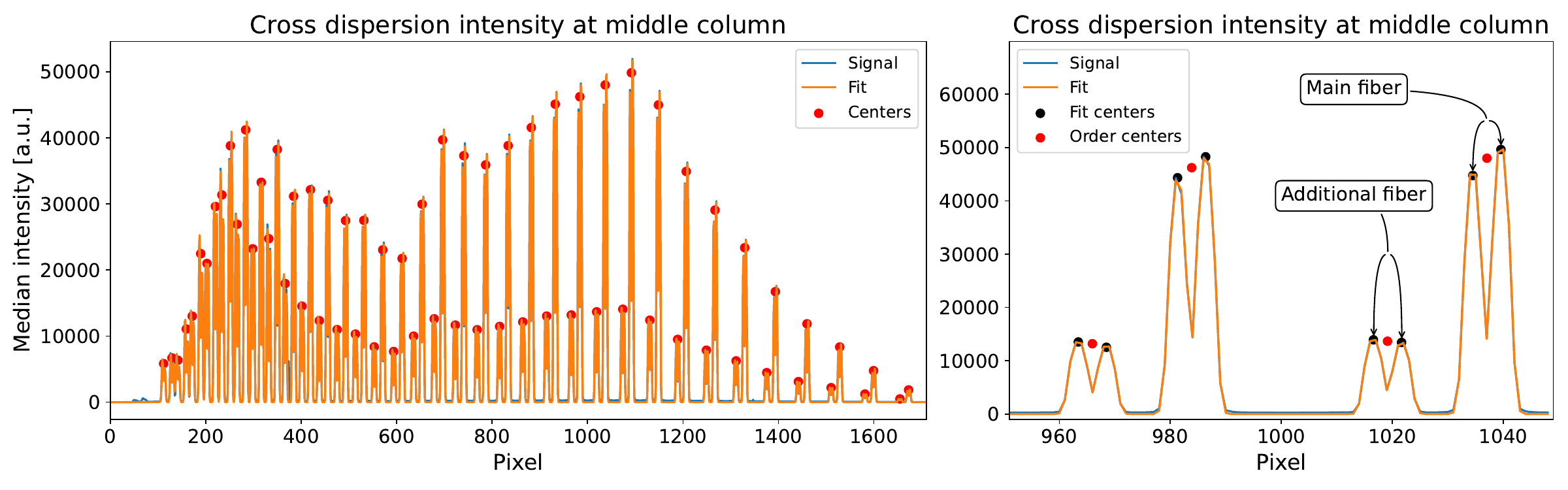}
\caption{Order detection for the FEROS spectrograph. This spectrograph has both an image slicer and two independent optical fibers, which are imaged side by side on the detector. LEFT: Cross dispersion cut through the orderdef image at the central column. The signal is shown in blue, while the fit is displayed in orange. Both agree to a large extend. Each peak highlighted with an orange dot in this image defines the position of one diffraction order. RIGHT: Zoom into the left image. The centers of both components of the double Gaussian fit functions are shown as well as the center of the order. Both optical fibers of the spectrograph can be easily distinguished.\label{fig:order_detection}}
\end{figure*}

Before the spectrum can be extracted from the science frames, generally applicable CCD corrections must be performed first. A master bias is created from all available bias frames by calculating the median pixel by pixel. Similarly, the flat frames, the orderdef frames (if available), and the dark frames (if available) for each exposure time are combined. The master bias is subtracted from the master flat, master orderdef, and master dark frames. The flat correction differs from the approach used in photometry and is described separately in Sec. \ref{subsec:structure_flatcorr}. The exposures can be transposed, if necessary, so that the order traces run approximately along the pixel rows of the image. 

In addition, the effects of cosmic rays are filtered out of the science frames using the astroscrappy package \citep{Dokkum2001, McCully2018}. Since the master bias, master dark, master flat, and master orderdef are derived from the median of the individual exposures, the effects of cosmic rays are already have been removed for those types of frames. In the ThAr images, cosmic rays cannot be filtered out, as the algorithm could result in filtering out some of the desired ThAr emission lines. Cosmic rays may be identified as false emission lines there. However, due to the typically short exposure time of ThAr frames --- which results in few cosmic rays --- and the large number of ThAr emission lines, the effect on wavelength calibration is negligible.

\subsection{Order detection} \label{subsec:structure_orderdetection}
\begin{figure*}[t!]
\centering
\includegraphics[width = \textwidth]{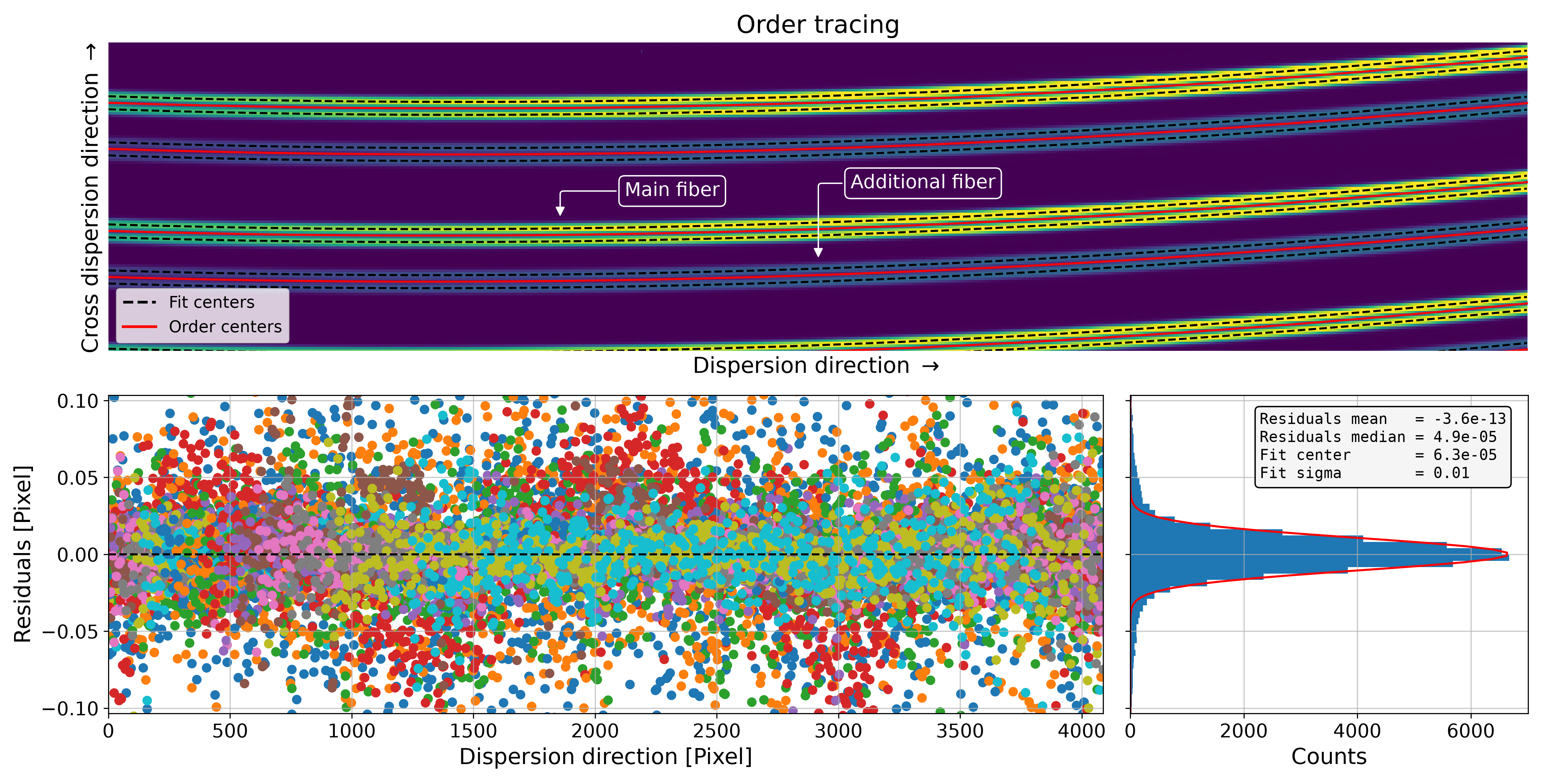}
\caption{Order tracing for FEROS. TOP: Excerpt of the orderdef frame of FEROS with the traced order positions. The paths of the fit centers and the paths of the order centers are shown. BOTTOM LEFT: Residuals between the fit and the measured values. All orders are plotted. The worst 1\% of the residuals have been excluded. BOTTOM RIGHT: Histogram of the residuals and Gaussian fit. \label{fig:order_residuals}}
\end{figure*}
To extract the spectra, the position of the diffraction orders on the detector chip must be known. Orderdef frames are used for this task. First, the diffraction orders are searched for along the central column of the image, as this is where most of the signal is present. To minimize outliers, not only is the signal from the central column used, but the median is also calculated using the pixels from the neighboring columns. By default, the width of this band is five pixels, although this value can be adjusted depending on the spectrograph. This results in the signal being approximately perpendicular to the diffraction orders, i.e. along the cross dispersion direction. Peaks are then searched for in the 1D signal. These peaks determine the positions of the orders in this column, as can be seen in the left part of Fig. \ref{fig:order_detection}. To determine the position of the orders with subpixel accuracy, the signal is fitted with different functions, depending on the optical design of the spectrograph, as shown in the right part of Fig. \ref{fig:order_detection}. If the spectrograph does not have an image slicer, the orders are fitted with a single modified Gaussian function:
\begin{equation}
S(x) = A \cdot e^{-0.5 \cdot \left(\frac{\left|x - c \right|}{\sigma} \right)^p}
\end{equation}
Here, $S(x)$ is the signal at pixel $x$, $A$ is the amplitude, and $c$ is the center of the function. $\sigma$ determines the width of the function, and $p$ is a generalized exponent. If the spectrograph has an image slicer, a double modified Gaussian is fitted instead of a single modified Gaussian function:
\begin{equation}
S(x) = A _1\cdot e^{-0.5 \cdot \left(\frac{\left|x - c_1 \right|}{\sigma_1} \right)^{p_1}} + A_2 \cdot e^{-0.5 \cdot \left(\frac{\left|x - c_2 \right|}{\sigma_2} \right)^{p_2}}
\end{equation}
For a spectrograph without an image slicer the center of the order is $c$. In case of an image slicer, the mean of the two centers $c_1$ and $c_2$ is used. If the spectrograph has multiple optical fibers, the orders are sorted into the individual fibers. Individual order detections that cannot be clearly assigned to a specific fiber are filtered out, as they are very likely false-positives.

Each detected order is now tracked individually across the detector. Starting from the central column, the detection moves to the right in a specific step size. Since the position of the order does not change drastically, it is not necessary to determine the position for each column; instead, only every fifth column is taken into account. Once enough positions of an order have been determined, the centers of the order are iteratively fitted with a global polynomial. The degree of this polynomial must be determined manually for each spectrograph. A fifth-degree polynomial has proven to be a good starting point. As soon as enough order centers were determined to reliably fit the order trace polynomial, this polynomial is used to estimate the expected position of the order at the next measurement point. This allows the fit of the center of the order to be performed with significantly greater precision. If there are not yet enough data points for the iterative fit of the order trace polynomial, the position of the last measurement is used as the starting point for the center fit. When the right edge of the image is reached, the procedure continues from the center to the left. The position of the orders is ultimately determined by the global polynomial fit of all measured positions. If the signal at the edges of the detector becomes too weak to continue measuring the order center, the order trace is extrapolated using the global polynomial. At those points, the order trace is no longer well-defined. As the order trace polynomial is significantly overdetermined by the large number of measurements and a low polynomial degree, the order trace continues without significant deviations even in the extrapolated region. However, since there is no signal present anyway, any minor deviations are irrelevant. A detail of an orderdef exposure from the FEROS spectrograph can be seen in the upper part of Fig. \ref{fig:order_residuals}. Since the position determination of the individual orders is independent of each other, it is processed in parallel to increase speed. After order tracing, the individual orders are analysed with regard to their mean curvature, and outliers are filtered out, as they are very likely detection errors. Orders that do not lie entirely on the detector --- that is, extend above or below the detector --- are also rejected.

\begin{figure*}[t!]
\centering
\includegraphics[width =  \textwidth]{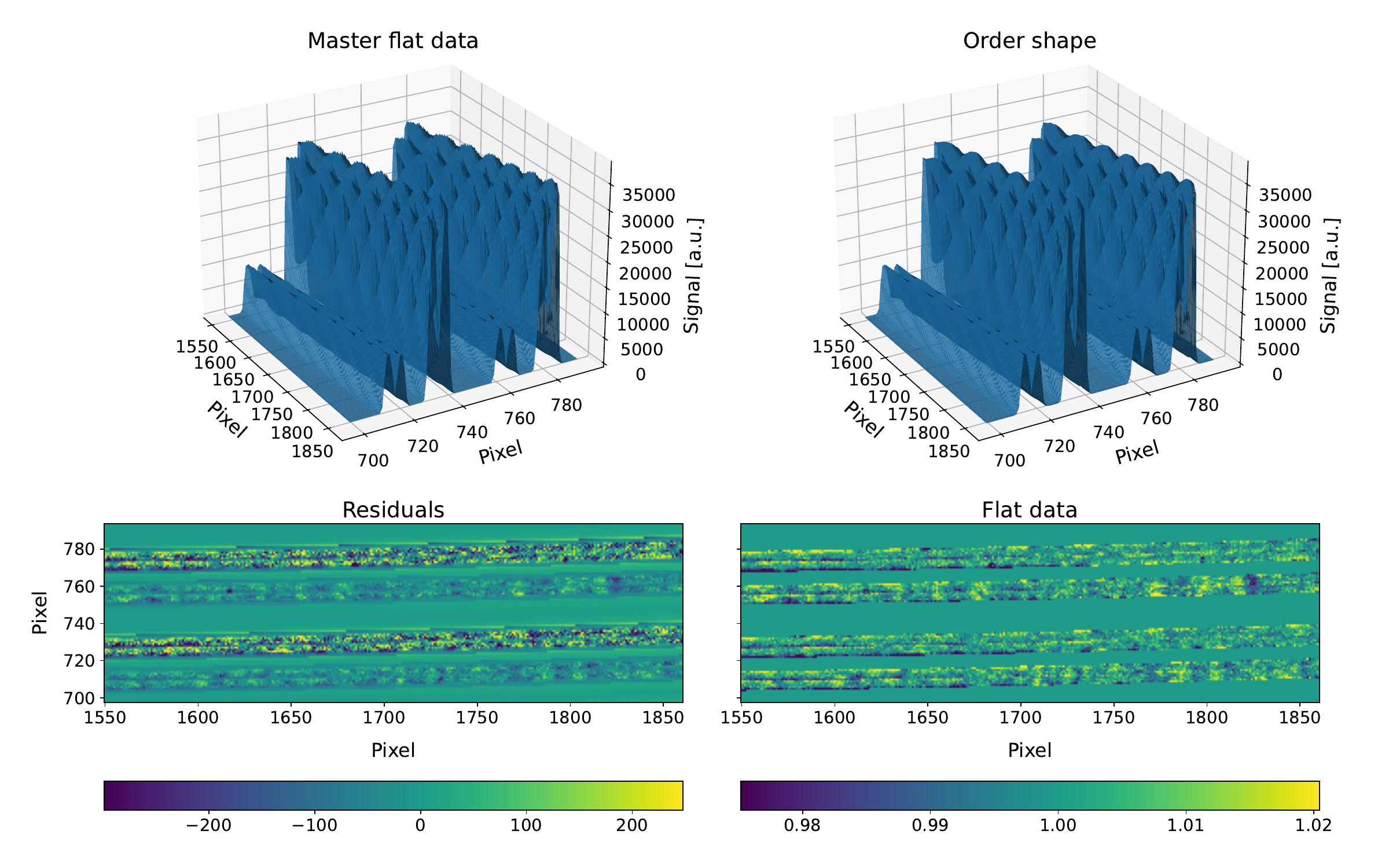}
\caption{Illustration of how the flat image is generated, using FEROS as an example. The figure shows a section of two diffraction orders, each with both the main fiber and the additional fiber. TOP LEFT: Section of the raw data. The fluctuations in the raw data are caused by the blaze function. TOP RIGHT: Section of the column order shape image. The fluctuations in the blaze function are also mapped. The differences from the raw data result solely from the varying sensitivities of the individual pixels. BOTTOM LEFT: Plot of the residuals between raw data and column order shape. BOTTOM RIGHT: Plot of the flat image. \label{fig:Flat}}
\end{figure*}

Even if the signal approaches zero at the edges, the complete diffraction order is still always extracted from edge to edge. For users who work with data from individual orders rather than the combined spectrum, the spectrum header specifies, for each individual order, the minimum and maximum pixels at which measurements of the order position could still be made. The position of the diffraction order is extrapolated for all pixels below and above these threshold pixels.

The order traces follow the measured order centers very closely. As can be seen in the lower part of Fig. \ref{fig:order_residuals}, the standard deviation for FEROS across all orders is only about $0.01$ pixels. For HARPS and the Shelyak eShel II, the standard deviations of the residuals are similarly small as well, namely $0.0075$ pixels and $0.015$ pixels, respectively.

Fitting a symmetric function can introduce systematic errors when tracing the order position if the cross-dispersion profile is not symmetric. This is the case, for example, with HARPS. In practice, however, this effect is smaller than one pixel, so the order remains well within the extraction window. The symmetric profile is also used exclusively for determining the order positions. For the subsequent determination of the order shape (see Sec. \ref{subsec:structure_ordershape}) a more general form is assumed, which may be asymmetric.

To distinguish between the individual orders, each detected diffraction order is assigned an order number $k$, where the first order is given the number $k \, = \, 0$ and each subsequent order has $k$ increased by one. This order number does not correspond to the physical diffraction order number, which is determined during wavelength calibration (see Sec. \ref{subsec:wavelength_thoughts}).

\subsection{Flat correction} \label{subsec:structure_flatcorr}
Similar to photometry, a flat-field correction is used to correct for the varying quantum efficiencies of the individual pixels. In photometry, the detector is illuminated as uniformly as possible to detect the differences between the individual pixels. Some spectrographs, e.g., ESPRESSO\footnote{see ESPRESSO Pipeline User Manual, Page 23} and UVES\footnote{see UVES User Manual, Page 46}, have a built-in flat lamp for this purpose, which illuminates the detector directly. For most instruments, however --- especially the more affordable ones --- this is not the case. For those instruments, the detector would have to be removed from the instrument, which is impractical. Instead, light from a source that is as spectroscopically flat as possible is directed into the spectrograph. 

In flat-field correction in photometry, the image to be corrected is divided pixel-by-pixel by the flat-field image. However, since a large portion of the pixels in spectroscopic images are not illuminated, this leads to a significant amplification of noise in these regions, as the division is performed by values close to zero. Rather, a method is used that is similar to the one described in  \cite{Piskunov2002}.

Instead of dividing the image directly by the flat image during correction, the measured flat spectrum (see upper left panel of Fig. \ref{fig:Flat}) is compared to a model (see upper right panel of Fig. \ref{fig:Flat}). Since this model describes the shape of the order along one pixel column, we also call it column order shape. The relative quantum efficiency of each pixel is determined by comparing the pixel value with the column order shape. The correction image generated in this way has values close to 1 at the locations of the diffraction orders that characterize the deviations in quantum efficiency, and values exactly equal to 1 for all values outside the orders (see lower right panel of Fig. \ref{fig:Flat}).

The flat image is calculated only for pixels that have a sufficiently high SNR and do not deviate too much from the column order shape. For pixels with too low SNR, the calculation of the flat image is too heavily influenced by noise. Excessive deviations indicate errors other than a slight difference in the quantum efficiency of the pixels (such as a hot pixel). However, these errors are not corrected as part of the flat-field correction, so the flat value for these pixels is set to 1. The master flat or master orderdef (if available) image is used for the calculation. Since these images are created from the median of several individual exposures, effects such as cosmic rays are already minimized. An example of how the flat image is generated can be seen in Fig. \ref{fig:Flat}.

The calculation of the column order shape is described in the following section. To calculate the column order shape, the positions of the diffraction orders on the detector are required. The flat-field correction can therefore only be performed after order detection (see Sec. \ref{subsec:structure_orderdetection}).

\subsection{Order shape calculation} \label{subsec:structure_ordershape}
Once the positions of the diffraction orders on the detector are known, a model of the shape of the orders can be derived. When determining the order traces, it was assumed that the shape of the orders corresponded to a modified Gaussian.  Although this model works well for determining the position, it does not correspond to the actual shape, particularly at the edges.

The form of the column order shape can vary along an order. For an accurate determination, the underlying illumination function $f(x, u)$ must be determined. This illumination function specifies the subpixel-accurate representation of the intensity distribution in the cross-dispersion direction. The actual measured values are the integral of this function over the pixels. We assume that the pixels are square and have no gaps in-between, which is the case for modern detectors. The noise-free column order shape $G(x,y)$ of the pixel at position $(x, y)$ is then given by
\begin{equation}
G(x,y) = \int_{y - y_{\mathrm{center}} - 0.5}^{y - y_{\mathrm{center}} + 0.5} f(x,u) \, \mathrm{d}u
\end{equation}

\begin{figure*}[t!]
\centering
\includegraphics[width =  \textwidth]{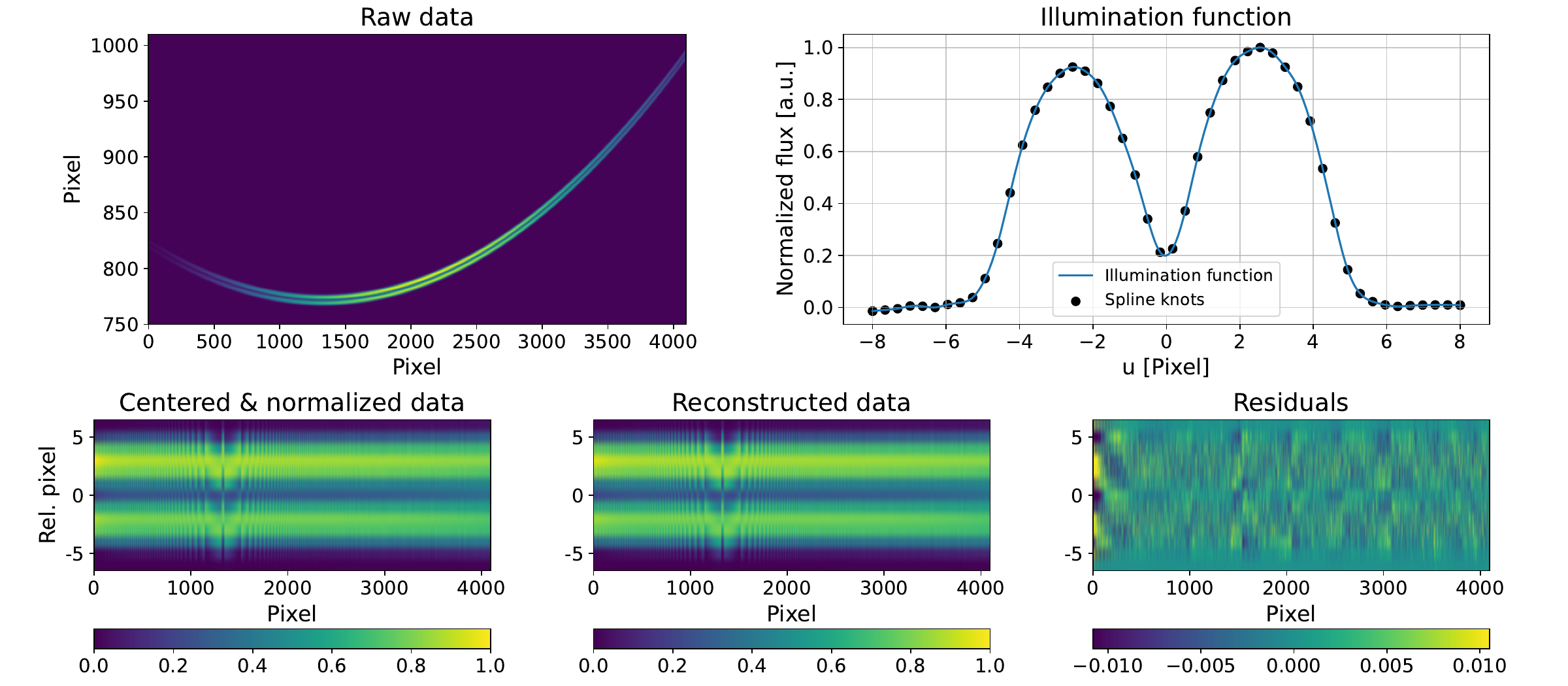}
\caption{Calculation of the column order shape. TOP LEFT: Raw data from a diffraction order of FEROS on the camera chip.  TOP RIGHT: Plot of the illumination function $f(x,u)$ for pixel $x = 2000$. The individual spline knots are also shown. BOTTOM LEFT: The raw data has been normalized and centered. The data was shifted so that the center of the order is now located within the middle pixel. The visible patterns are the subpixel shifts of the order relative to the pixel grid. BOTTOM CENTER: Reconstructed signal based on the illumination function. This image has the same scale as the raw data image. BOTTOM RIGHT: Residuals between raw and reconstructed data. \label{fig:Ordershape}}
\end{figure*}

Similar to the description in \cite{Piskunov2021}, the illumination function $f(x,u)$ must be known with subpixel accuracy. This takes advantage of the fact that, due to the curvature of the diffraction order on the detector, there are various offsets of the illumination function relative to the pixel grid. PyReduce divides the individual diffraction orders into several swaths and assumes that $f(x,u)$ is approximately constant within each swath and therefore does not depend on $x$. We use a more general approach by explicitly retaining the dependence on $x$. Therefore, there is no reason here to divide the diffraction order into individual swaths. 

We define the illumination function using a two-dimensional spline function consisting of cubic splines. Let $A_i(x)$ and $B_j(u)$ be B-spline functions in the $x$ and $u$ directions, respectively; then the illumination function is given by
\begin{equation}
f(x,u) = \sum_{i=0}^{n_x} \sum_{j=0}^{n_u} c_{i,j} \cdot A_i(x) \cdot B_j(u) 
\end{equation}
with the fit parameters $c_{i,j}$. The values $n_x$ and $n_u$ represent the number of spline knots in the $x$- and $u$-directions, respectively. An example of such an illumination function for a specific pixel column $x$ is shown in the upper right part of Fig. \ref{fig:Ordershape}. The fit should be able to accurately represent the illumination function, which is why $n_u$ is set to a high value. A value of $n_u = 50$ has proven to be suitable for instruments with an image slicer, and $n_u = 30$ for instruments without an image slicer. It should be noted here that this results in multiple knots per pixel. The function $f(x,u)$ can still be reconstructed because, due to the varying shift of the diffraction order relative to the pixel grid, there are enough measurements to reconstruct the illumination function with subpixel accuracy. The number of knots in the $x$-direction, on the other hand, should be kept small, since the shape of the illumination function changes only slowly. A value of $n_x = 10$ has become the standard here.

To ensure that the underlying function $f(x,u)$ --- and not the integratel $G(x,y)$ --- is fitted, we can take advantage of the linearity of the integral:
{\small
\begin{equation}
\begin{split}
G(x,y) &= \int_{y - y_{\mathrm{center}} - 0.5}^{y - y_{\mathrm{center}} + 0.5} f(x,u) \, \mathrm{d}u \\
&= \int_{y - y_{\mathrm{center}} - 0.5}^{y - y_{\mathrm{center}} + 0.5} \sum_i \sum_j c_{i,j} \cdot A_i(x) \cdot B_j(u) \, \mathrm{d}u \\
&= \sum_i \sum_j c_{i,j} \cdot A_i(x) \cdot \int_{y - y_{\mathrm{center}} - 0.5}^{y - y_{\mathrm{center}} + 0.5} B_j(u) \, \mathrm{d}u \\
&= \sum_i \sum_j c_{i,j} \cdot A_i(x) \cdot U_j(y)
\end{split}
\end{equation}
}

\begin{figure*}[t!]
\centering
\includegraphics[width =  \textwidth]{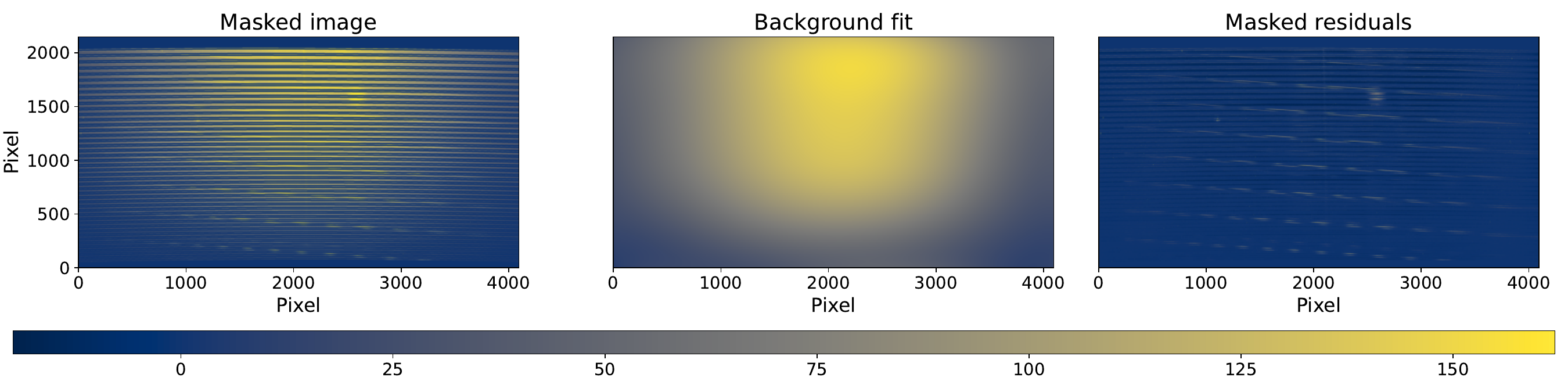}
\caption{Example of background subtraction for a spectrum from HARPS' blue component. All images are on the same scale. LEFT: Bias-corrected image. The detected orders were masked, and the corresponding pixels were set to zero. A large band of the image was masked above the first and below the last detected order to minimize the influence of any unrecognized orders at the edges of the image. CENTER: Background fit with $degx \,= \, degy \, = \, 8$. RIGHT: Residuals between the image and the background. The detected orders were masked again. It is evident that the background fit removed the large-scale trend, but smaller effects remain visible. \label{fig:Background}}
\end{figure*}

Since the integral depends only on $u$, only the integral over the $B_j(u)$ needs to be calculated. However, since the $B_j(u)$ are B-splines, these functions are already predetermined given the knot positions; thus, the integral $U_j(y) = \int_{y - y_{\mathrm{center}} - 0.5}^{y - y_{\mathrm{center}} + 0.5} B_j(u) \, \mathrm{d}u$ can also be precomputed. Therefore, despite the integral nature of $G(x,y)$, the fit of the parameters $c_{i,j}$ is a simple least-squares fit.

No interpolation is necessary to determine the illumination function. The only information that must be known is the position of the center of the order relative to the pixel grid obtained from order tracing. In order to speed up the calculation, the raw data is shifted so that the center of each row is aligned with the middle pixel (see upper left and lower left part of Fig. \ref{fig:Ordershape}).  In addition, the raw data for each pixel column is normalized to a sum of one. When normalizing the individual pixel columns, it has become apparent that simply dividing by the sum is not sufficient. The sum is highly sensitive to outliers in individual pixels. Due to the small number of pixels along the columns, the median is also subject to significant fluctuations. A stable approach has proven to be smoothing the sums of all pixel columns using a running median filter and then normalizing them using this median.

Using the normalized and centered data, we can now fit the illumination function $f(x,u)$. This requires solving the following minimization problem:
\begin{equation}
\begin{split}
&\sum_x \sum_y w_{x,y} \left[ E\left(x, y\right) - G\left(x, y\right) \right]^2 + \\
&\Lambda_x \cdot \sum_x \sum_k \left[ \frac{\partial^2 f(x,u)}{\partial x^2} \left(x_i, u_k\right) \right]^2 + \\
&\Lambda_u \cdot \sum_x \sum_k \left[ \frac{\partial^2 f(x,u)}{\partial u^2} \left(x_i, u_k\right) \right]^2 + \\ 
&\Lambda_b \cdot \sum_x \sum_b \left[ W_{b} \cdot \left| f\left(x_i, u_b\right) \right| \right] \\
&= \mathrm{min}
\end{split}
\end{equation}

Here, $E(x,y)$ is the measured signal at pixel $(x,y)$, and $w_{x,y}$ are the weights of the individual pixels. The first part of the equation represents the standard minimization problem of a least-squares fit. The remaining terms are used to better constrain the shape of $f(x,u)$. $\Lambda_x$ and $\Lambda_u$ weight the second derivative of $f(x,u)$ in the $x$ and $u$ directions, respectively. The larger these values are, the flatter the function becomes. This allows unwanted oscillations to be dampened. $\Lambda_b$, on the other hand, weights the function values at the boundary points $u_b$. These boundary points can be flexibly chosen within a boundary region of the function and assigned weights $W_b$. This final term ensures that the function $f(x,u)$ drops to zero in the $u$-direction at the boundaries and does not produce any unphysical boundary effects.

Using the illumination function calculated in this way, the order shape for each pixel column $x$ can now be calculated. This result is used, for example, to calculate the flat image. The centered results and the residuals relative to the raw data are shown in the lower part of Fig. \ref{fig:Ordershape} in the center and on the right. Normalized to a profile peak of $1$, the RMS is approximately $0.0054$ for HARPS, approximately $0.0046$ for FEROS, and approximately $0.042$ for the Shelyak eShel II.

It should be noted here that the illumination function determined in this way --- and thus also the column order shape --- may be asymmetric. In contrast to order tracing (see Sec. \ref{subsec:structure_orderdetection}), an asymmetric form of the diffraction order --- such as that observed with HARPS --- is therefore fitted appropriately here.

\subsection{Background subtraction} \label{subsec:structure_background}
\begin{figure*}[t!]
\centering
\includegraphics[width =  \textwidth]{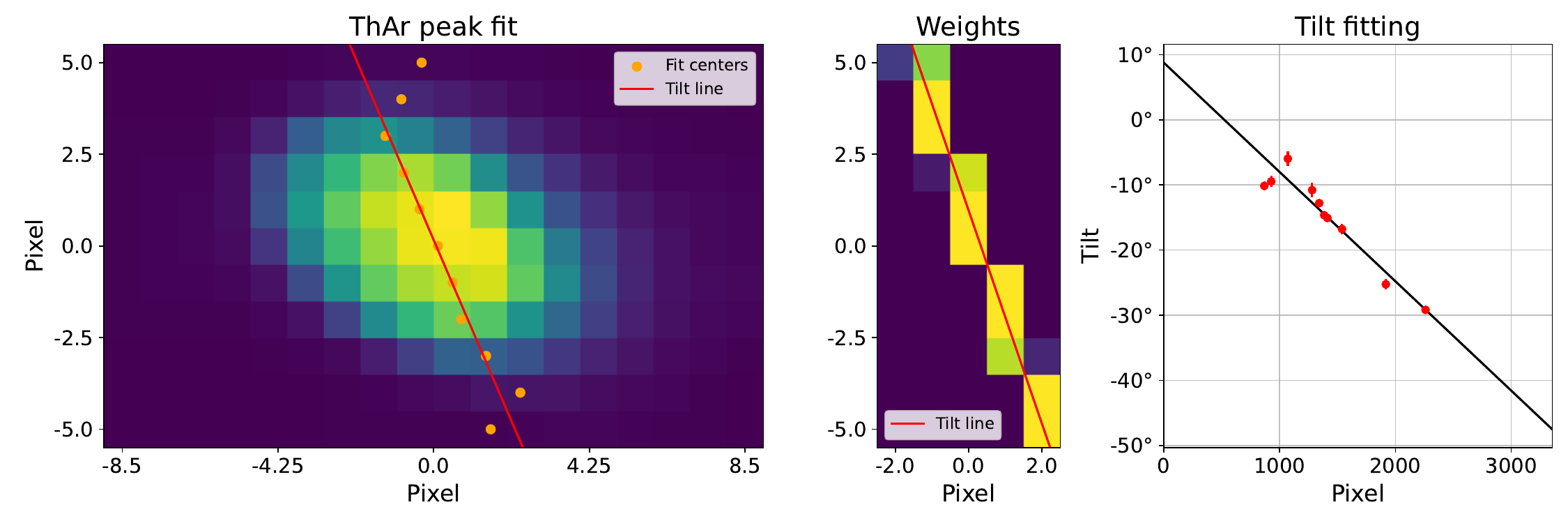}
\caption{Determination of the weights for the Shelyak eShel II spectrograph. This spectrograph exhibits a significant tilt in the emission lines, which must be taken into account during the extraction. LEFT: Determination of the tilt using a ThAr emission line. Each individual pixel row is fitted with a Gaussian function. The centers obtained from these functions are approximated by a straight line to calculate the tilt. CENTER: Weights derived solely from the tilt. The weights are no longer parallel to the pixel columns. RIGHT: Linear fit of the measured tilts over one order.  \label{fig:weights}} 
\end{figure*}

Not all light in the spectrograph reaches the detector directly; some light is scattered onto the sensor. This light is also called the background and must be removed. Since the orders and the scattered light overlap, the background must be interpolated based on the signal between the individual orders. Therefore, this correction is only possible after determining the positions of the orders, as the orders must be masked. 

The background is fitted using a two-dimensional Chebyshev polynomial. The polynomial degree in the $x$ (row) and $y$ (column) directions must be determined manually for each spectrograph. A starting point of $degx \,= \, degy \, = \, 8$ has proven to be effective. If there is significant background between the individual orders, resulting in a fit based on many data points, the polynomial degree can also be increased to allow for a better fit of the background. This is the case, for example, with the Shelyak eShel II. In the case of closely spaced orders, as with FEROS, the polynomial degree should be kept relatively low, since a large portion of the image is masked and high polynomial lines tend to cause unwanted oscillations in these areas. The background subtraction removes large-scale trends but not smaller effects such as ghosts, as shown in Fig. \ref{fig:Background}.

\vspace*{0.5cm}
\section{Spectrum extraction} \label{sec:spectrum_extraction}
PyVesta offers two different methods for extracting one-dimensional spectra from two-dimensional images. The first method is called ''very fast extraction'' and is described in Section \ref{subsec:extraction_veryfast}. It is primarily used for a quick overview and should not be used for scientific analysis.  

For accurate data extraction, the ''weighted extraction'' algorithm should be used. This algorithm is based on the optimal extraction routine described by \cite{Horne1986} and \cite{Marsh1989}. The algorithm was adapted to our specific case of tilted lines and is described in Section \ref{subsec:extraction_weighted}.

\subsection{Very fast extraction} \label{subsec:extraction_veryfast}
With this method, no weights are assigned to individual pixels, nor is the full width of the diffraction order used. Instead, based on the calculated position of the order, the pixel closest to the center of the order is chosen. The value of this pixel now represents the spectrum at this pixel column for the current order. If the order does not pass perfectly through the center of a pixel, the two nearest pixels are added, with the weighting increasing the closer the center of the pixel is to the center of the order. This calculation is performed for each column, resulting in the extraction of the spectrum.

This method is used to quickly obtain an initial overview of the spectrum. This overview can, for example, be displayed immediately after the acquisition. Above all, however, this extraction method is used to extract a ThAr spectrum in order to determine the approximate positions of the emission lines. These emission lines are required to calculate the weights for the ''weighted extraction'' method.

\subsection{Weighted extraction} \label{subsec:extraction_weighted}
This method is used to extract scientifically useful spectra. All pixels representing the signal of a given wavelength are weighted and summed to produce a final spectrum. For this extraction, it is important to determine which pixels correspond to the same wavelength and how they should be weighted. The process of determining which pixels correspond to a given wavelength is described in section \ref{subsubsec:extraction_weighted_linetilt}. The determination of the weighting is explained in section \ref{subsubsec:extraction_weighted_weightcalculation} and the final extraction algorithm in section \ref{subsubsec:extraction_weighted_extraction}.

\subsubsection{Line tilt determination} \label{subsubsec:extraction_weighted_linetilt}
To simplify, we assume that a single wavelength is mapped onto a straight line within a specific order on the detector. However, this line does not have to align with the pixel columns; instead, it may be tilted relative to them. We refer to the angle between the pixel columns and this line as the tilt angle. A tilt of $0^\circ$ therefore corresponds to an orientation along the pixel columns. For spectrographs with excellent optical correction, such as HARPS and FEROS, this tilt is negligible, so the determination of the weights for these spectrographs can be simplified. However, the Shelyak eShel II, an off-the-shelf spectrograph that is significantly less expensive than the aformentioned spectrographs, exhibits a noticable tilt \citep{Stock2025}. This tilt angle is not constant but varies between individual diffraction orders and even within a single order. A light source with infinitely sharp emission lines would be ideal for determining the tilt angle. However, such a source is not physically possible; nevertheless, the ThAr exposures required for wavelength calibration provide images with sufficiently sharp  emission lines. To identify these emission lines, a ThAr spectrum extracted by the ''very fast extraction'' method is used. From the lines identified in this way, those that either have a neighboring line too close by or are too bright and overexpose the detector are filtered out, since the tilt cannot be clearly determined for these lines. 

The tilt for each line is determined individually. Each pixel row of the line is fitted with a Gaussian, and its center is determined. The centers of these individual rows are then fitted with a straight line using a weighting that corresponds to the square of the SNR (signal to noise ratio). An example of such a fit is shown in the left panel of Fig. \ref{fig:weights}.

The measured tilts are then fitted by a Chebyshev polynomial to calculate tilts for all pixels without an emission line. The user can decide whether to perform a fit of each order individually or a 2D fit across all orders simultaneously. For spectrographs with significant vignetting near the edges of the order, only low polynomial degrees should be used, as the polynomials otherwise tend to exhibit excessive fluctuations at the edges. In the case of our Shelyak eShel II, linear fits have proven to be suitable. Such a fit is shown in the right panel of Fig. \ref{fig:weights}.

\begin{figure*}[t!]
\centering
\includegraphics[width =  \textwidth]{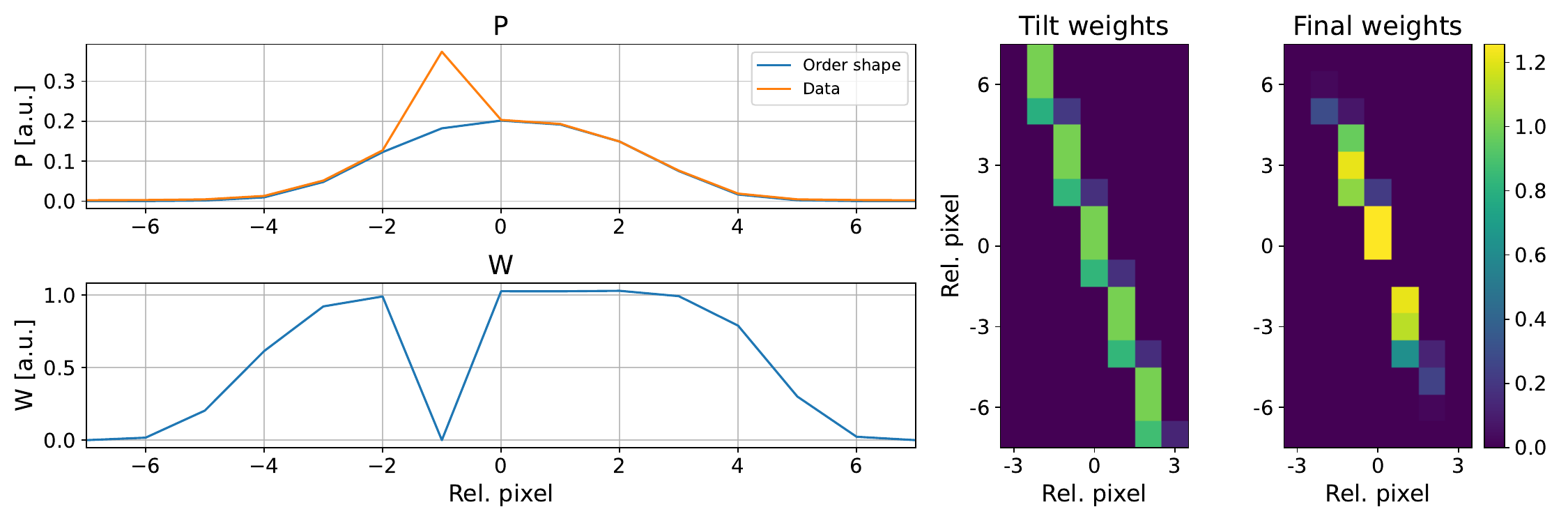}
\caption{Example of determining the final extraction weights. TOP LEFT: Example of a cross-dispersion distribution along one wavelength and the corresponding wavelength order shape for reference. It is clear that one of the pixels is affected by an error. BOTTOM LEFT: Weights W. The defective pixel was excluded from the weights. The pixels at the edges, which contribute less signal, have lower weights, too. RIGHT: Tilt weights as described in Sec. \ref{subsubsec:extraction_weighted_weightcalculation} and final weights. To calculate the final weights, the tilt weights are multiplied by W row by row. To ensure that the total flow is conserved despite the exclusion of a single pixel, the remaining pixels are weighted more heavily. Both images are on the same scale. \label{fig:Extraction}} 
\end{figure*}

\subsubsection{Weights calculation} \label{subsubsec:extraction_weighted_weightcalculation}
The weights for the extraction algorithm determine how strongly individual pixels contribute to the extracted spectrum. The weights are represented as a two-dimensional array, with the center of the weights coinciding with the intersection of the order trace and the current pixel column. In the case of a negligibly small tilt, the weights correspond to a single pixel column, where each pixel is assigned a weight of 1.

If the tilt is not negligible, the adjacent pixel columns must also be taken into account, since the tilt causes a wavelength to no longer be mapped along a single column. A line is then generated across the weights' array based on the tilt. The greater the portion of a pixel that lies on this line, the higher the weight assigned to that pixel. Pixels that do not touch the line have a weight of zero. We refer to the weights generated in this way as tilt weights. The sum of the weights in a row of pixels is always equal to 1. An example of such tilt weights is shown in the center panel of Fig. \ref{fig:weights}.

Since the calculation of the weights is independent of one another, this process can be easily divided into separate threads to increase processing speed. They are saved in a separate file along with the corresponding pixel boundaries, which define where the weights are located in the image. Given that the tilts are assumed to remain constant throughout the night, these tilt weights are used for all exposures, regardless of whether they are science or calibration frames. However, for the final extraction, the SNR of the individual pixels must still be taken into account, as described in the following section.

\subsubsection{Extraction} \label{subsubsec:extraction_weighted_extraction}
The extraction process is based on the principle of ``optimal extraction`` as described by \cite{Horne1986} and \cite{Marsh1989}. It is assumed that the flux $F$ is distributed among different pixels $i$. Each pixel $i$ receives the proportion $P_i$ of the total flux. Neglecting noise, each pixel should therefore measure the signal
\begin{equation}
D_i = P_i \cdot F.
\end{equation}
The total signal is simply given by
\begin{equation}
F = \sum_i D_i,
\end{equation}
since $\sum_i P_i = 1$. In this simple case, however, all pixels are added together regardless of their contribution to the signal. By weighting the pixels according to their contribution the error of the spectrum can be minimized. Using the weights $W_i$, the spectrum is now given by 
\begin{equation}
F = \sum_i W_i \cdot D_i.
\end{equation}
\cite{Horne1986} has shown that, for minimal variance, the weights $W_i$ are
\begin{equation}
W_i = \frac{P_i / V_i}{\sum_i P_i^2 / V_i},
\label{eq:Weights}
\end{equation}
where $V_i$ is the variance of the individual pixels. For all pixels whose error is dominated by the Poisson noise of the photon distribution, $V_i \sim D_i \sim P_i$ holds, and thus $W_i \sim \text{const}$. For pixels where the detector's readout noise has a significant effect, the weights decrease.

Determining the weights $P_i$ is very important for this algorithm. Here, the approach differs from that of \cite{Horne1986} and \cite{Marsh1989}. While both assume that one wavelength is aligned along the pixel columns, this is no longer the case in our approach. 

The $P_i$ define the shape of the order in the cross-dispersion direction and are thus similar to the column order shape described in Sec. \ref{subsec:structure_ordershape}. The wavelength of the individual pixels was irrelevant for calculating the flat image, which is why the column order shape was determined there along the pixel columns. For the extraction, however, a different order shape is required, one that takes into account the tilt of the emission lines. Since the order shape is now calculated for one wavelength, we'll call it wavelength order shape for clarity. When generating the tilt weights, a fit of the wavelength order shape is therefore calculated using the master flat by determining the order profile in the cross-dispersion direction along a single wavelength rather than along the pixel columns. Otherwise, the procedure is very similar to determining the column order shape in Sec. \ref{subsec:structure_ordershape}. This wavelength order shape only needs to be generated once, since the shape of the orders does not change overnight. Although the wavelength order shape can change over the course of the night, just as the wavelength solution can, both effects are in the subpixel range. While this shift is relevant for the wavelength solution, it is negligible for the wavelength order shape.

The $P_i$ values are thus defined directly by the wavelength order shape and do not need to be determined for each science image. However, depending on the image's SNR, the weights of the individual pixels will still vary. The final weights are therefore determined individually for each image according to equation (\ref{eq:Weights}).

Since the form of the wavelength order shape is constant, it can also be used to detect individual defective pixels. In Fig. \ref{fig:Extraction}, for example, it is clearly visible that one pixel is producing an excessively strong signal. This pixel can be identified by its deviation from the wavelength order shape and can therefore be masked. The masked weights $W^m_i$ correspond to the original weights $W_i$, but the weights of all masked pixels are set to 0. However, masking a pixel causes the total signal to decrease, as it is the sum of all pixels. To compensate for this, all other weights are increased so that the flux is conserved. The wavelength order shape is used for this, as it can be assumed to be free of defects. The new weights $W'_i$ are calculated from the masked weights $W^m_i$ and a factor $s$
\begin{equation}
W'_i = s \cdot W^m_i,
\end{equation}
where $s$ is determined by the wavelength order shape $P_i$, the masked weights $W^m_i$, and the original weights $W_i$
\begin{equation}
s = \frac{\sum_i W_i \cdot P_i}{\sum_i W^m_i \cdot P_i}.
\end{equation}
The weights $W'_i$ are finally multiplied row by row by the tilt weights, as can be seen in the right-hand side of Fig. \ref{fig:Extraction}.

For the extraction, the section of the spectrum corresponding to the current weights is now multiplied pixel by pixel by the final weights and eventually summed. This process is performed individually for each pixel of every diffraction order. Since the individual orders are assumed to be independent, they are extracted in parallel across multiple processes to save time. It should be noted, however, that the extraction process is primarily memory-limited; therefore, running too many parallel processes can actually slow down the extraction rather than speed it up.

\section{Wavelength calibration} \label{sec:wavelength_calibration}
\begin{figure*}[t!]
\centering
\includegraphics[width = \textwidth]{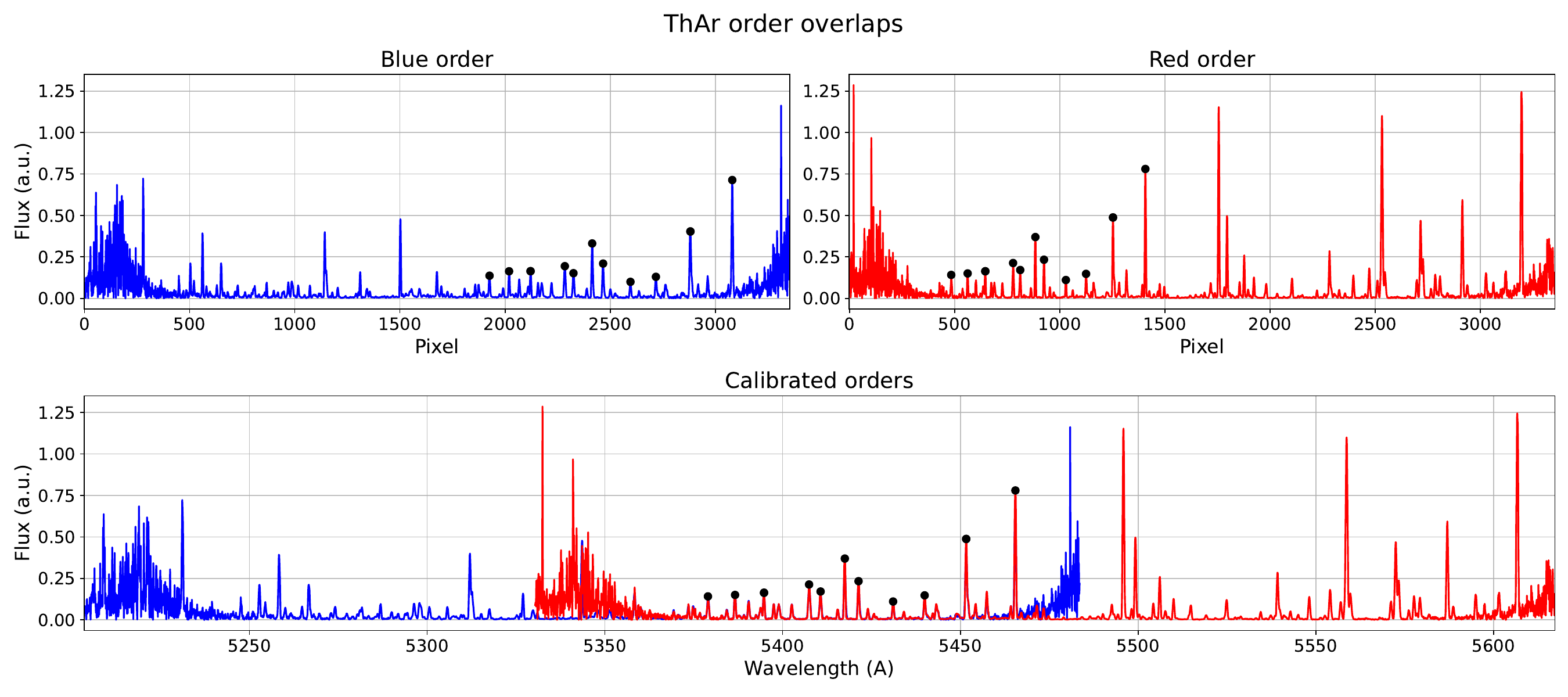}
\caption{TOP: ThAr emission spectrum of physical diffraction orders 41 (red) and 42 (blue) of the Shelyak eShel II spectrograph. ThAr emission lines detected in both orders are marked with black dots. Based on these detected emission lines, the general course of the wavelength resolution can be determined. BOTTOM: The ThAr emission spectra of the two orders overlayed. \label{fig:Overlaps}}
\end{figure*}

\subsection{General thoughts on wavelength calibration} \label{subsec:wavelength_thoughts}
The extracted spectra are currently defined only in the pixel domain. However, in order to scientifically analyze the spectra, the signal must be known in terms of the light’s wavelength. The conversion between the pixel and wavelength scales is called wavelength solution. The vast majority of spectrographs are based on the physical principle of a reflection grating. According to the grating equation, the diffraction maxima of this grating are given by
\begin{equation}
m \cdot \lambda = d \cdot \left( \sin \alpha + \sin \beta \right),
\end{equation}
where $m$ is the physical diffraction order, $\lambda$ is the wavelength, $d$ is the grating constant, and $\alpha$ and $\beta$ are the angles of the incident and emerging light beams, respectively. Since $\alpha$ and $\beta$ in particular are generally unknown, the wavelength is approximated using a polynomial \citep{Brahm2017}: 
\begin{equation}
\lambda(x,m) = \frac{1}{m} \sum_{i=0}^{n_m} \sum_{j=0}^{n_x} a_{ij} c^{i}(m) c^{j}(x)
\end{equation}
Here, $\lambda$ is again the wavelength of pixel $x$ in diffraction order $m$. $a_{ij}$ are the coefficients of the polynomial, and the $c^{n}$ are Chebyshev polynomials of degree $n$.

However, the physical order $m$ does not usually correspond to the index of the found order $k$. Instead, we define a value $m_0$ such that 
\begin{equation}
m = m_0 + k.
\end{equation}
$m_0$ corresponds to the physical diffraction order of the first detected order. This value varies for each spectrograph and must be determined before starting the wavelength calibration.

\begin{figure*}[t!]
\centering
\includegraphics[width = \textwidth]{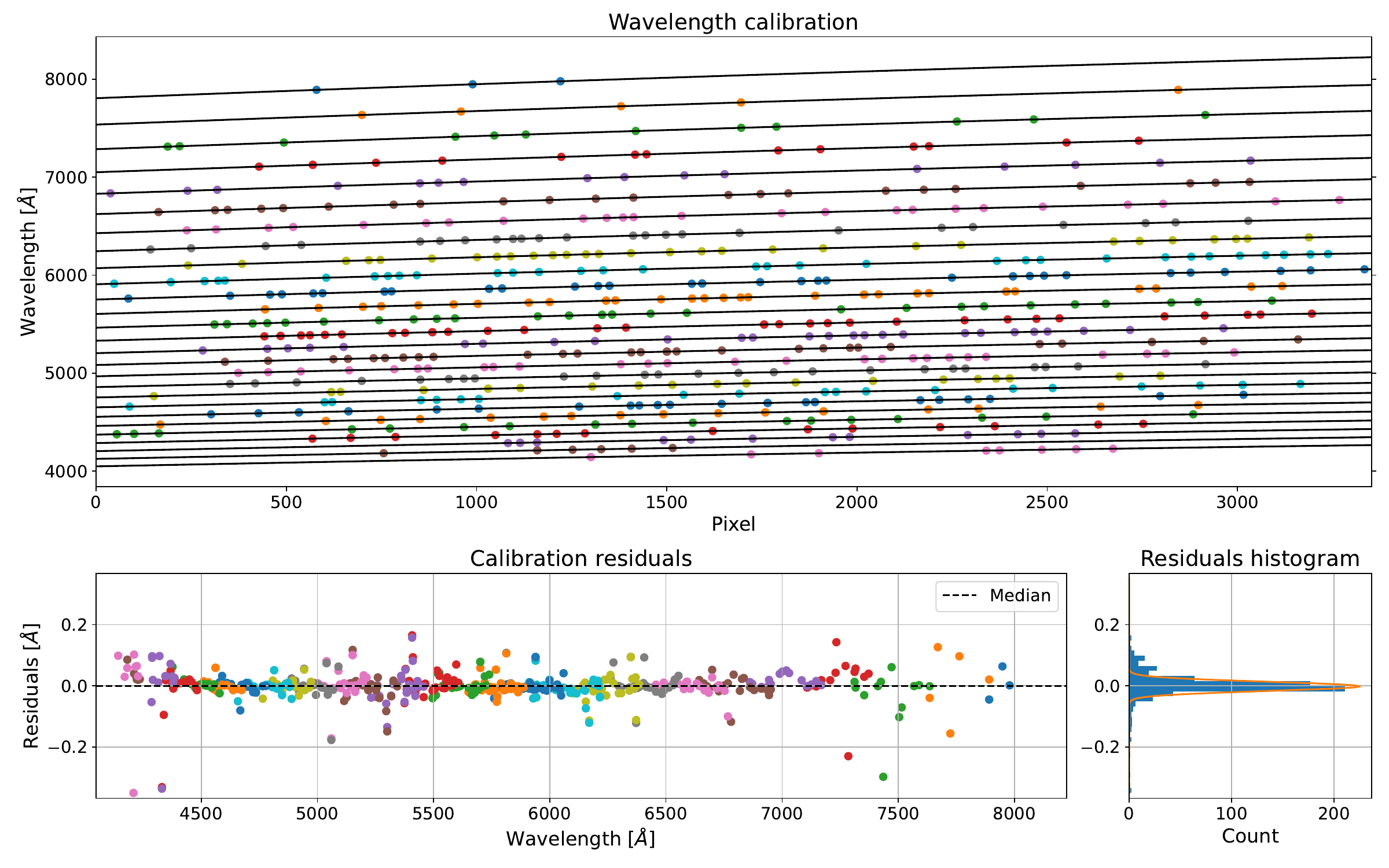}
\caption{TOP: Wavelength solution for the Shelyak eShel II spectrograph. Each line represents the wavelength of a diffraction order. The points are the ThAr emission lines used for wavelength calibration. BOTTOM LEFT: Residuals relative to the reference for the individual ThAr emission lines. BOTTOM RIGHT: Histogram of the residuals. The histogram was fitted using a Gaussian to determine the error of the wavelength solution. \label{fig:WavelengthCalibration}}
\end{figure*}

\vspace*{0.5cm}
\subsection{Different types of wavelength calibration} \label{subsec:wavelength_differenttypes}
To determine the wavelength at a specific point in the spectrum, it is important to have a reference spectrum whose pattern can be used to pinpoint the exact position in the spectrum. Such reference spectra are typically generated as emission spectra. Therefore, substances are vaporized in a gas discharge lamp. Various substances, such as Hydrogen (H), Helium (He), or Mercury (Hg), are suitable for this purpose. For the wavelength calibration of high-resolution spectrographs, it has become standard practice to use a gas discharge lamp containing the elements Thorium (Th) and Argon (Ar). This combination provides an emission spectrum with many sharp emission lines across the entire visible spectrum. For this emission spectrum, there are lists indicating at which wavelengths in the spectrum an emission line (hereinafter referred to as a ''peak'') is present. There are two ways to compare this list with the measured raw spectrum:

The first method relies on the fact that the spectrograph has already been characterized. This allows a list to be created that contains the pixel position and wavelength of the peaks for each diffraction order. Using this list, a simple cross-correlation between the list and the raw spectrum can be used to determine the correct order and any possible shift at the pixel level. From that, one can also derive $m_0$. This method is very fast and stable. The major drawback of this method, however, is that the reference list containing the pixel positions of the peaks must first be generated. This list is not available for all spectrographs and cannot be created without an additional method for determining the wavelengths. 

The second method is based on the fact that, in most spectrographs, there is spectral overlap between two consecutive diffraction orders. This method was first introduced by \cite{Brandt2020}. The implementation in this software closely follows their approach. Because of this overlap, the spectrum --- and thus the emission line pattern --- is present at two different positions on the detector. These different positions must therefore correspond to the same wavelength. Since we now have two positions for a wavelength, we can infer the shape of the wavelength solution from this. An example of those overlaps can be seen in Fig. \ref{fig:Overlaps}. Only a global scale factor cannot be determined solely based on the overlaps. However, this scale factor can be calculated by comparing the determined wavelengths with a list of the reference wavelengths of the peaks. Unlike the previous method, it is sufficient here to have a general list of all reference wavelengths; one does not need to know their exact pixel positions. Such lists depend only on the emitting materials and are readily available for ThAr lamps. A more detailed description of this method can be found in  \cite{Brandt2020}.

\begin{figure*}[t]
\centering
\includegraphics[width = \textwidth]{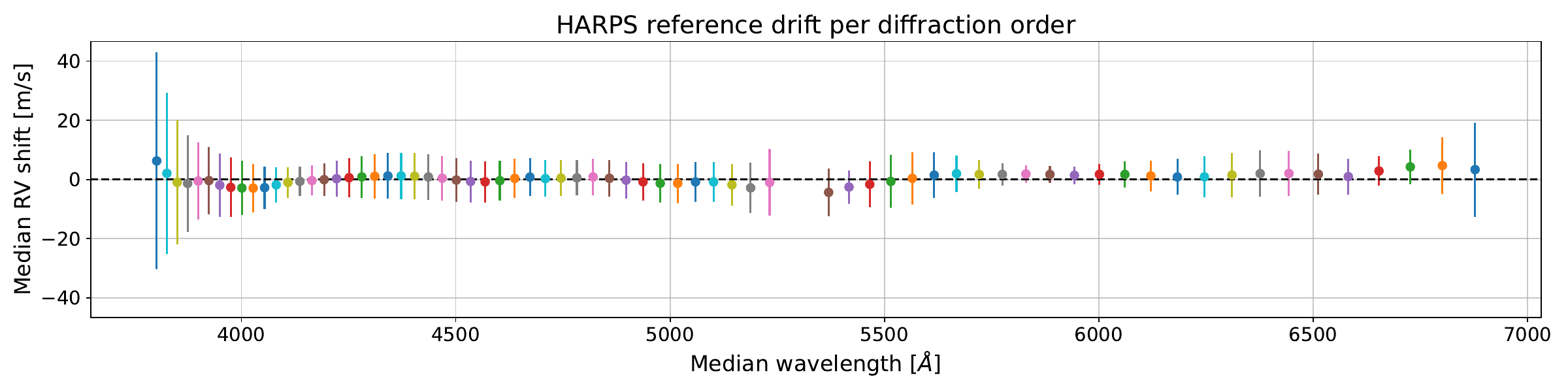}
\caption{Illustration of the wavelength drift for the individual diffraction orders of the HARPS spectrograph. A wavelength solution from the beginning of the observation night is compared to a wavelength solution from the end of the observation night. The error bars correspond to the RMS of the residuals between the wavelength solutions. Note that the error in the wavelength solution of approximately $26.9$ m/s is not taken into account. The median shift is approximately $-0.2$ m/s in the blue component and $1.3$ m/s in the red component. A weak pattern is noticeable in the wavelength shifts, but it is significantly smaller than the accuracy of the wavelength solution.
\label{fig:HARPS_wavedrift}}
\end{figure*}

Since the second method is better suited for use with various spectrographs, this software uses the determination of wavelengths via overlaps. A wavelength solution of the Shelyak eShel II spectrograph can be seen in Fig. \ref{fig:WavelengthCalibration}. 

However, for spectrographs for which a reference list of ThAr peaks is already available, a wavelength solution can also be generated based on this list. In this case, this method is preferable, as it is usually faster and more stable. PyVesta includes sample code demonstrating how to create a reference list using a wavelength solution generated via overlaps. 

Some spectrographs, such as HARPS, offer wavelength calibration using ThAr lamps as well as wavelength determination using a laser frequency comb (LFC) or Fabry-P\'erot etalons (FP). Since LFCs can be adjusted with great precision, they can be used to determine a very accurate wavelength based on a rough preliminary wavelength estimate. However, since maximum wavelength accuracy is not the goal of this software and off-the-shelf spectrographs do not use LFCs, PyVesta does not support wavelength determination using LFCs. FPs are not used to determine the wavelength, but rather to detect small changes in the wavelength solution over the course of the night. Therefore, they are usually imaged onto a second fiber alongside the starlight to ensure the highest accuracy. Since wavelength interpolation is sufficiently accurate for our current purposes (see the following section), PyVesta does not yet support shift determination using a FP as well. Other calibration options, such as an iodine cell, are also currently not supported.  

\subsection{Wavelength solution stability} \label{subsec:wavelenth_stability}
\begin{figure*}[t!]
\centering
\includegraphics[width = \textwidth]{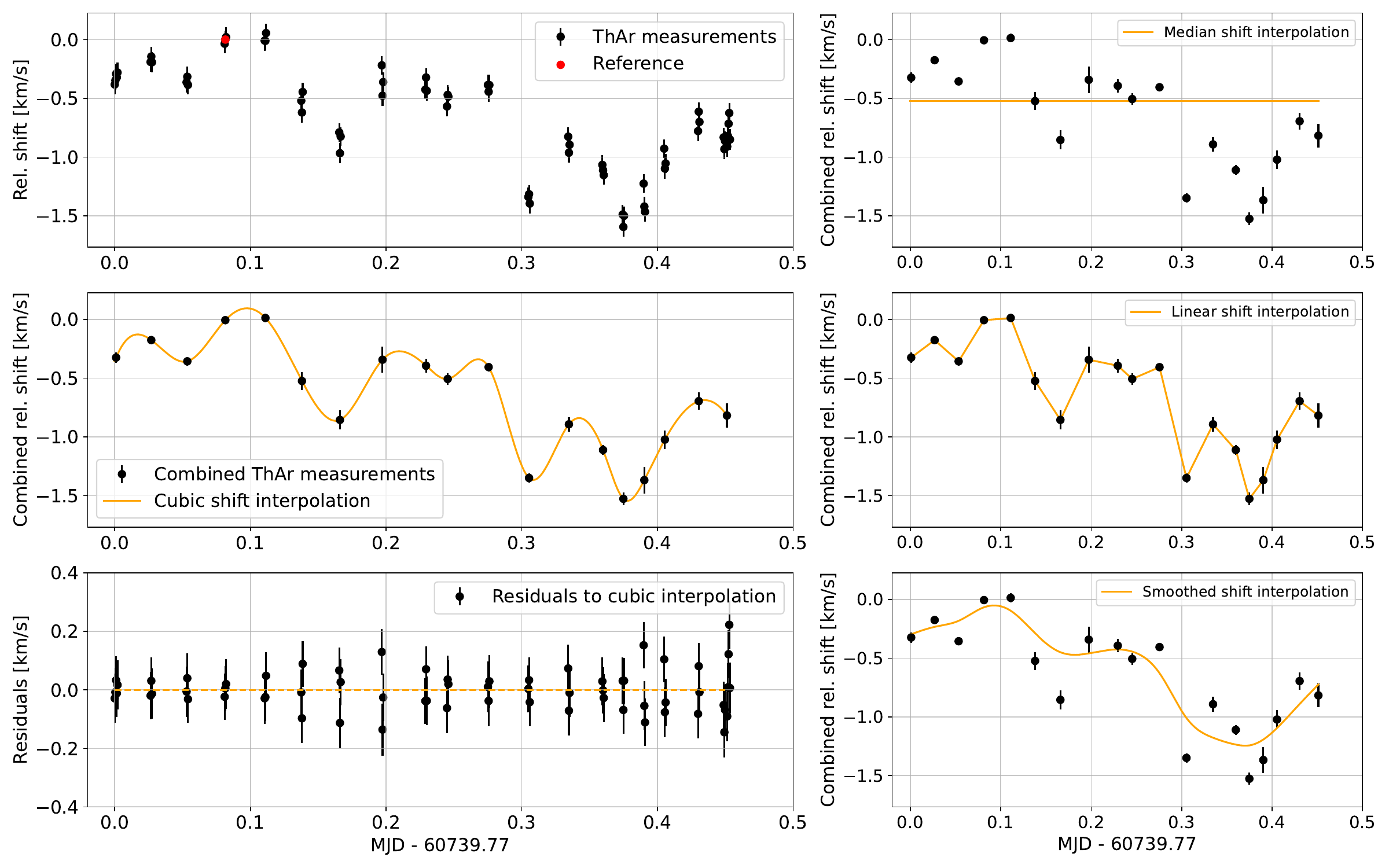}
\caption{Different methods of interpolating the wavelength solution shift of the Shelyak eShel II spectrograph.  TOP LEFT: All shifts of the ThAr frames from a single night. At least 3 ThAr frames were taken in immediate succession. It can be seen that these consecutive frames always show almost the same shift. The wavelength solution with the lowest RMS was used as a reference and is marked here in red. CENTER LEFT: All related ThAr frames were combined to reduce the shift error. The shift is now interpolated using a cubic spline. BOTTOM LEFT: Residuals of the individual ThAr frames for interpolation. \\
RIGHT: Various other types of interpolation. The type of interpolation should be adjusted depending on how often and when ThAr frames are acquired. TOP RIGHT: Median interpolation, uses only the median of all shifts. CENTER RIGHT: Linear interpolation, interpolation using a linear spline. BOTTOM RIGHT: Smoothed interpolation, interpolation using a smoothing spline.
\label{fig:WavelengthInterpolation}}
\end{figure*}

\begin{figure*}[t!]
\centering
\includegraphics[width = \textwidth]{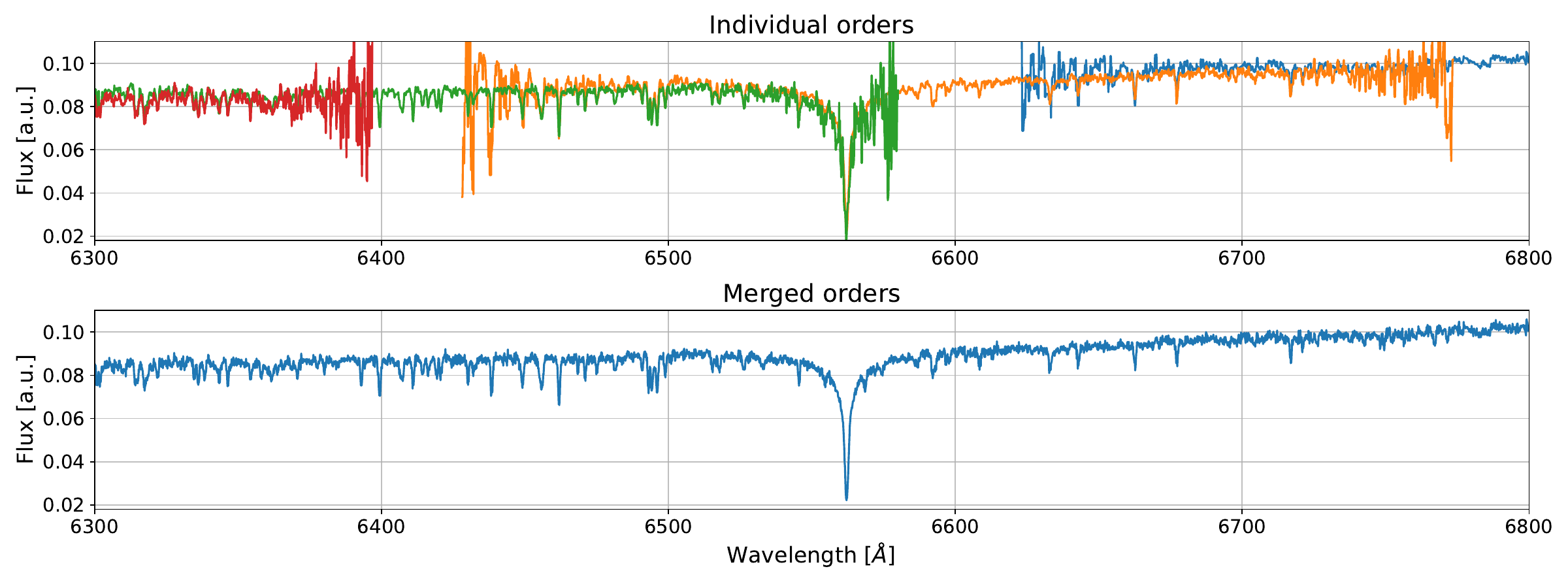}
\caption{TOP: Spectra of multiple diffraction orders from the Shelyak eShel II spectrograph of the star $\tau$ Boo. It is clearly visible that the orders overlap (similar to the ThAr images). Note that the noise increases significantly at the edges of the orders, since there is almost no signal there due to the spectrograph’s blaze function. BOTTOM: Merged spectrum of $\tau$ Boo. The noise from the individual orders is no longer visible. \label{fig:merging}}
\end{figure*}

During a night of observations, several ThAr frames are typically acquired for wavelength calibration. A separate wavelength solution is generated for each of these frames. However, spectrographs might not be perfectly stable over observation time, so the individual wavelength solutions differ from one another. On one hand, changes in temperature and air pressure may cause mechanical changes in the spectrograph’s geometry. Vibrations caused by motions of the telescope or the dome can also affect the spectrograph. For high-resolution spectrographs such as HARPS and FEROS, this is not a major problem, as they are operated in specially stabilized rooms located away from the telescope \citep{Mayor2003, Kaufer1999}. Smaller instruments, however, are usually located directly next to the telescope, so fluctuations in temperature and air pressure, as well as disturbances caused by the motion of the telescope, cannot be avoided \citep{Stock2025}. While the stabilized spectrographs thus exhibit significantly fewer absolute deviations, their resolution is higher than that of smaller instruments, so compensating for the spectrograph’s instability is advisable in all cases.

The instability of a spectrograph is described, using the Doppler effect, as a radial velocity shift $\Delta v$ between the observed wavelength solution and the reference solution:
\begin{equation}
\lambda = \left(1 + \frac{\Delta v}{c} \right) \cdot \lambda_{\mathrm{ref}}
\end{equation}
Here, $\lambda$ and $\lambda_{\text{ref}}$ are the wavelengths of the solution under consideration and the reference solution, respectively, and $c$ is the speed of light in vacuum. By definition, the shift is defined separately for each wavelength. It is therefore possible that the shift differs for each wavelength. In addition to the RV change, a translational movement of the detector may also occur, which likewise alters the wavelength solution. At constant resolution, this translation effectively results in an RV shift. If the spectrograph’s resolution is not constant, however, a wavelength-dependent shift may be observed. In \cite{Stock2025} it was shown that the shift is, at least for our Shelyak eShel II, indeed independent of wavelength. We were also unable to detect any wavelength dependence of the shift for HARPS and FEROS, which was greater than the error of the wavelength solution. A per-order comparison of the wavelength solutions from HARPS at the beginning and end of a night of observations is shown in Fig. \ref{fig:HARPS_wavedrift}. Therefore, we use only the RV shift to determine the instrument’s stability and changes in the wavelength solution over the course of the night. For very high-resolution observations, the wavelength solution should be determined more precisely, but that is not the purpose of this software.

We investigated the source of the Shelyak eShel II's instability. The findings revealed that fluctuations in temperature or air pressure are not the main causes; rather, the mechanical movement of the telescope --- and in particular the dome --- is responsible for changes in wavelength resolution. A more detailed description of this can be found in \cite{Stock2025}.
 
We therefore interpolate the instrument’s shift by interpolating the various measurements throughout an observation. In most cases, not just a single ThAr calibration frame is acquired, but several ThAr frames in quick succession, which are combined to increase the accuracy of the measurement. Then these combined measurements are interpolated using a cubic spline. An example of this interpolation is shown in Fig. \ref{fig:WavelengthInterpolation}. In addition to cubic interpolation, linear interpolation, a smoothing interpolation or simply the median of the measurements can also be used. This should be determined for each spectrograph based on the number of ThAr images available. If ThAr images are available at regular intervals throughout the measurement, cubic interpolation is recommended. If ThAr images were only taken at the beginning and end of the observation, linear interpolation or the median is a better choice, as cubic interpolation can cause undesirably large fluctuations in this case. 

Thus, the shift relative to the reference wavelength solution can be determined for every moment of the night. The shift is corrected for each image, thereby significantly increasing the accuracy of the spectrograph. A significant improvement in the accuracy of radial velocity measurements using the Shelyak eShel II was demonstrated in \cite{Stock2025}. 

\section{Science frame processing} \label{sec:science_processing}
\subsection{CCD corrections \& Extraction} \label{subsec:science_CCD_extraction}
The raw data is corrected for bias and dark frames, if available. In addition, the background is subtracted and the effects of cosmic rays are removed. The ''weighted extraction method'' is always used to extract science images.

\subsection{Blaze correction} \label{subsec:science_blaze}
Flat-field correction for a spectral image is not performed in the same way as in photometry. Instead, the spectral sensitivity of the spectrograph must be corrected. Therefore, a lamp with as little spectral variation as possible is used for the flat-field images. The extracted science spectrum is then divided pixel-by-pixel by the extracted master flat spectrum. To distinguish it from the flat correction used in photometry, this is called the blaze correction.

\begin{figure*}[t!]
\centering
\includegraphics[width = \textwidth]{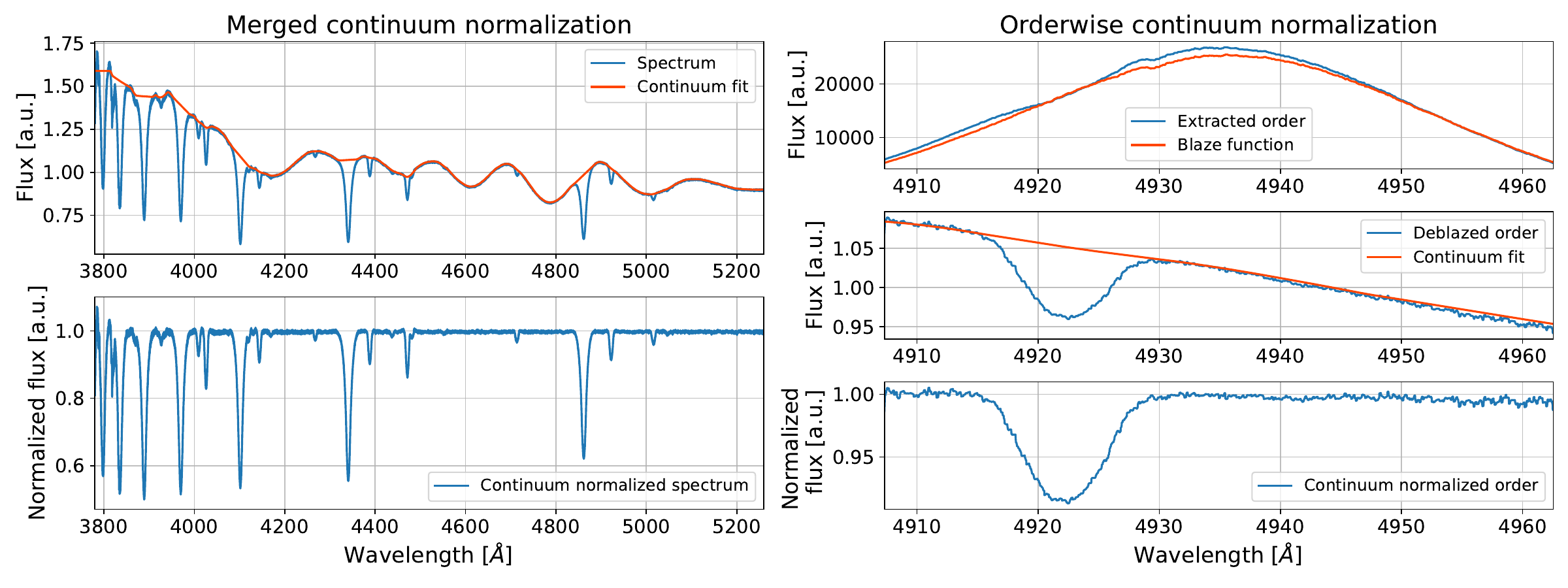}
\caption{TOP LEFT: Deblazed and merged spectrum of the star HD108257, recorded using the blue component of HARPS. It is clearly visible that, despite deblazing of the individual diffraction orders, the spectrum still shows artifacts resulting from the spectrograph’s sensitivity distribution. The continuum fit is also shown. The Balmer lines were masked for the continuum fit. The fit follows the continuum profile well, but some absorption lines influence the fit. For an analysis of these lines, it is better to use the unnormalized spectrum. BOTTOM LEFT: Normalized spectrum of HD108257. \\
TOP RIGHT: A single spectral order and the Blaze function for that order. CENTER RIGHT: Deblazed spectrum and continuum fit for that order. The continuum fit is the part of the continuum of the entire spectrum that has been cropped to match the wavelength range of the order. BOTTOM RIGHT: Continuum normalized order.  \label{fig:normalization}}
\end{figure*}

It is important to note, however, that while the flat spectrum ideally has no spectral features, in practice it will never exhibit a constant spectral intensity across the entire spectral range. Blaze correction therefore cannot correct the optical properties of the instrument entirely, as illustrated on the right side of Fig. \ref{fig:normalization}. A spectrum of a spectroscopic standard star would be necessary to calibrate the wavelength dependent response function of the instrument. Since the continuum of the stellar spectrum is normalized at the end by PyVesta anyway (see Sec. \ref{subsec:science_continuum}), no focus was placed on this. 

\subsection{Wavelength calibration} \label{subsec:science_wave}
After the wavelength solution has been determined, it is now applied to the spectrum. The mean time of the observation is used as the reference time for interpolating the shift. However, to perform accurate spectral analyses, particularly radial velocity measurements, the instrument’s relative velocity and light travel time to the barycenter of the solar system must also be calculated and corrected. For calculating the barycentric correction, the astropy package \citep{AstroPyCollaboration2022} is used, resulting in a numeric accuracy in the range of $3 \, \text{cm/s}$. 
 
\subsection{Order merging} \label{subsec:science_merging}
So far, the spectrum is still separated into its individual diffraction orders. This is the best approach for high-resolution spectroscopy and radial velocity measurements, as it allows the wavelength solution to be applied directly. However, for stars with very broad spectral lines,  to gain an overview of the spectrum or to be able to correlate multiple spectral lines at once, it may also be useful to combine all orders of the spectrum into a single continuous spectrum, as shown in Fig. \ref{fig:merging}. 

When combining the orders, several possibilities may arise. It might be that there is no spectrum at all for certain spectral regions. One example of this is the HARPS spectrograph, which is divided into a red and a blue section. However, there is a range between the two parts of the spectrograph that is not detected by either sensor. In this case, the software also skips this range. Additionally, there may be spectral ranges that occur in only one order. These should not be altered by the merging process in order to avoid compromising the accuracy of the wavelength calibration.

If a spectral range is mapped across two or more diffraction orders, the measurements from these orders must be combined. All measurements must first be interpolated to the same wavelength data points. The order with the highest SNR in this spectral range is used as a reference. This ensures that we use the wavelengths that are best defined, since the wavelength solution is poorest at the edges, where there is hardly any signal present. Interpolation to the new wavelengths is performed using linear interpolation of the original data. The merged flux is now calculated pixel-wise from the SNR-weighted average of the individual orders.

The implementation of the algorithms in this software is based on the methods implemented in the ''spectrum'' Python package \citep{Cokelaer2017}.

\begin{figure*}[t!]
\centering
\includegraphics[width = \textwidth]{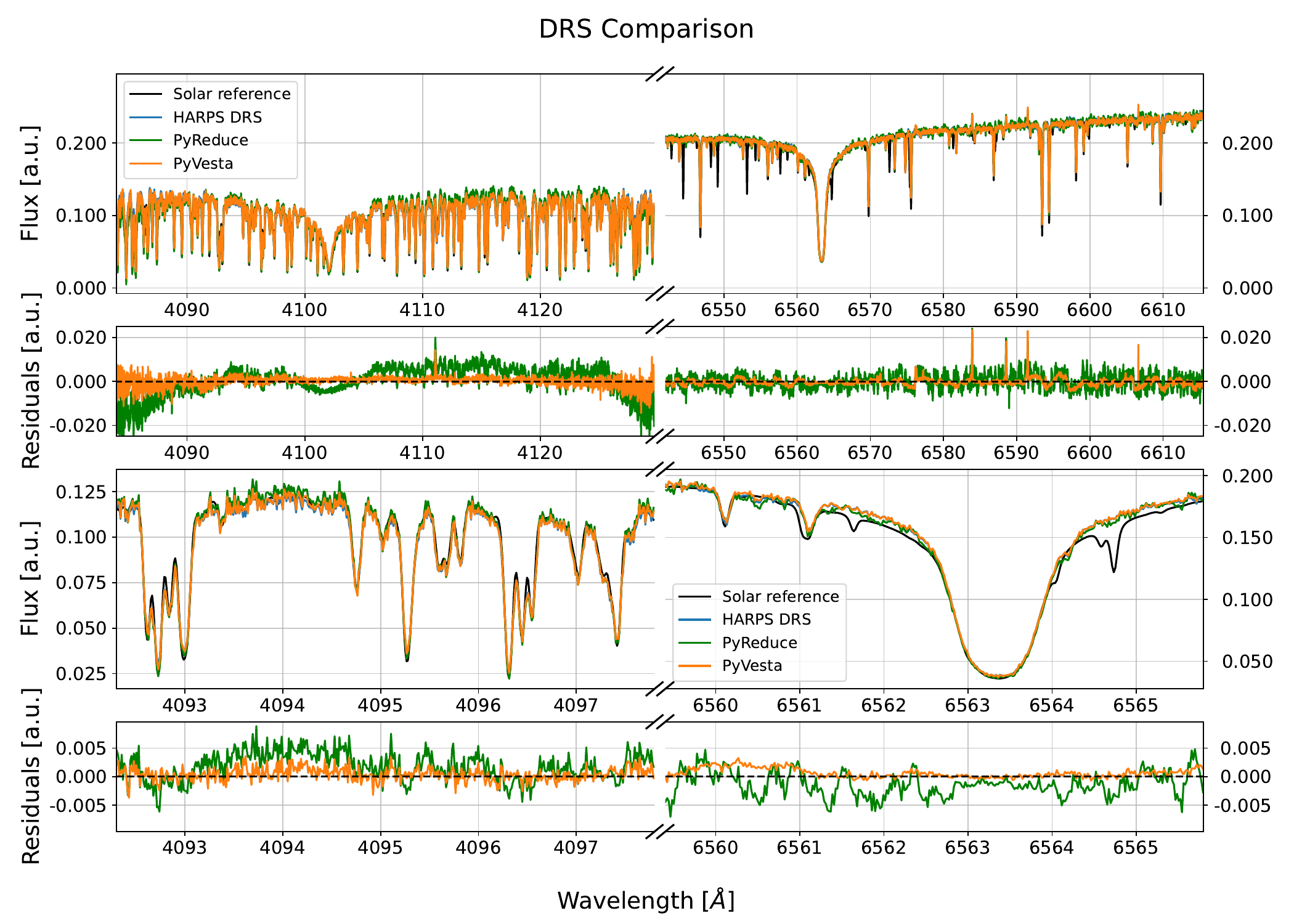}
\caption{Comparison between HARPS Standard DRS, PyReduce, and PyVesta. The figure shows a spectrum of the asteroid (4) Vesta --- that is, a reflected solar spectrum --- taken with HARPS, as well as a solar reference spectrum. Since the solar reference spectrum is normalized, the continuum was adjusted to match that of the HARPS observations. The features that do not match, particularly absorption lines around H$\alpha$, are tellurics that have been shifted by the barycentric correction and whose intensity depends heavily on the current state of the atmosphere. Excerpts from the blue component are shown on the left side, and excerpts from the red component are shown on the right side. The spectra were each scaled to the same median. All spectra were blaze-corrected using the master flat from PyVesta. The residuals are shown relative to the HARPS DRS, with the color coding preserved. A comparison with the reference spectrum would provide an objective comparison of all three pipelines, but residuals relative to the reference spectrum are dominated by the shifted tellurics, which makes it harder to see the differences between the individual pipelines. Wavelength calibration using PyReduce failed, so the wavelength solution from PyVesta is used instead. The spectrum shown is file \textit{HARPS.2011-12-01T00:23:12.542\_e2ds\_A.fits}, exposure time $600$ s, captured on December 1, 2011, with ESO programme ID 088.C-0323(A).\label{fig:Comparison}}
\end{figure*}

\subsection{Continuum normalization} \label{subsec:science_continuum}
The final step in processing the stellar spectra is the normalization of the continuum. As described in Sec. \ref{subsec:science_blaze}, no flux calibration is performed, and in many scientific applications --- especially for radial velocity measurements --- the spectral continuum is not relevant. Thus the spectral continuum is normalized to a constant value of one. 

In practice, it is very difficult to fit the continuum reliably. Depending on the spectral class of the target star, the spectrum may contain either a large number of narrow absorption lines or few very broad ones, as well as very strong emission lines, that should not affect the determination of the continuum. As described in Sec. \ref{subsec:science_blaze}, the merged spectra are still influenced by the spectral sensitivity of the spectrograph even after blaze correction. Accordingly, the shape of the apparent continuum also varies from spectrograph to spectrograph.  HARPS, for example, generates a wave pattern in the continuum of the observed stars; see Fig. \ref{fig:normalization} for details. This continuum is difficult to fit using polynomials, as these can lead to undesirable oscillations in the fit, particularly at the edges.

This software uses a method described in \cite{Piskunov2021}. The continuum is described by a function $f(x)$, where $x$ is the wavelength under consideration. The spectrum is given by $s(x)$, and each data point is weighted by the weights $\omega_x$. These weights are used, for example, to mask the regions where tellurics --- that is, absorption lines of the Earth’s atmosphere --- occur. Now, $f(x)$ is determined by solving the minimization problem \citep{Piskunov2021}:

\begin{equation}
\begin{split}
&\sum_i \omega_i \left[ f(x_i) - s(x_i) \right]^2 + \Lambda_1 \cdot \sum_i \left( \frac{\text{d} f}{\text{d}x} (x_i) \right)^2 + \\
&\Lambda_2 \cdot \sum_i \left( \frac{\text{d}^2 f}{\text{d}x^2} (x_i) \right)^2 = \text{min}
\end{split}
\end{equation}
Here, $\Lambda_1$ and $\Lambda_2$ are two numerical parameters that should be tested manually once for each spectrograph. The larger the value of $\Lambda_1$, the more constant the continuum fit becomes, since $\Lambda_1$ weights the derivative of the continuum. Similarly, a larger $\Lambda_2$ reduces small fluctuations, since it weights the second derivative of the continuum. 

In this way, the continuum can be fitted without the need of using a specific function. The continuum is thus defined by wavelength data points, not by a parametric function. Continuum normalization is then performed by simply dividing the star’s spectrum by the continuum pixel by pixel. 

The continuum can be determined either individually for each spectral order or using the merged spectrum. Both methods are supported, but calculating the continuum with the merged spectrum has proven to be more robust. For stars with very broad spectral lines, it is possible that the spectral line spans the entire order, meaning no continuum is available for the calculation. Instead, the continuum calculated from the merged spectrum can also be used to normalize the individual orders.

Continuum normalization may alter the shape of absorption lines. For a precise analysis, one should therefore use the unnormalized spectrum and perform a local continuum normalization.

\begin{figure*}[t!]
\centering
\includegraphics[width = \textwidth]{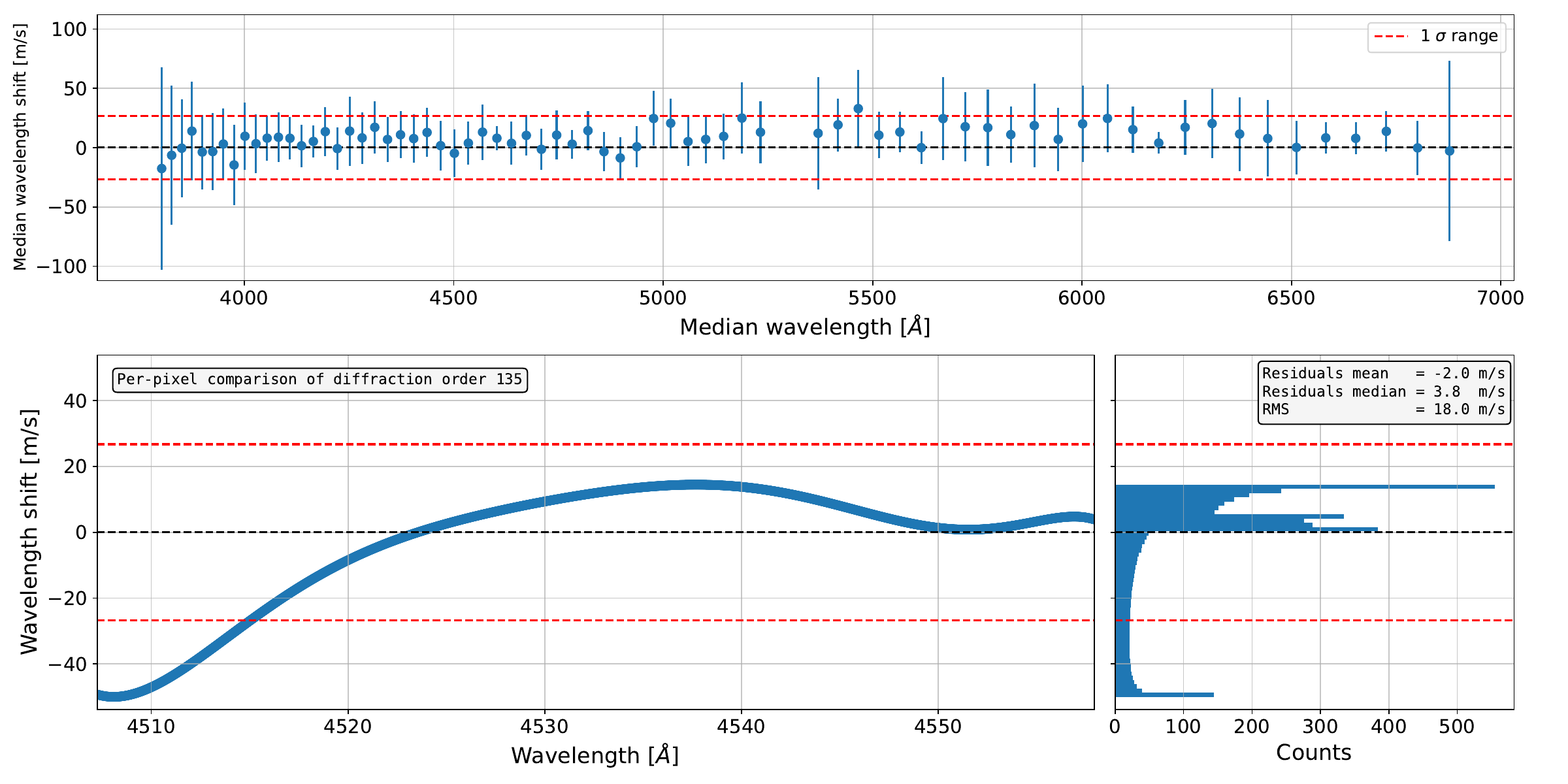}
\caption{Comparison between wavelength solutions from HARPS Standard DRS and PyVesta. TOP: Median shift per diffraction order. The error bars indicate the RMS between the two wavelength solutions and do not account for the calibration errors of the individual wavelength solutions. BOTTOM: Detail of the wavelength residuals for a single order. The wavelength solution is less well-defined at the edges because fewer calibration lines are available there due to the declining Blaze function. The histogram of the residuals is shown on the right. \label{fig:Comparison_wavesolutions}}
\end{figure*}

\section{Discussion \& Conclusion } \label{sec:conclusion}
\subsection{Comparison} \label{subsec:comparison}
As described in Section \ref{subsec:Discussion}, PyVesta is not designed to analyze individual instruments with the highest possible accuracy. Instead, PyVesta was designed for maximum compatibility so that new instruments can be easily added. Precisely because of this general-purpose design of PyVesta, it is interesting to compare PyVesta with a specialized data reduction software (DRS) and other similar programs. Fig. \ref{fig:Comparison} shows a spectrum of the asteroid (4) Vesta, analyzed using the HARPS Standard DRS, PyReduce, and PyVesta, based on data from both the blue and red components of HARPS. A solar reference spectrum \citep{Pietrow2026, Babcock1947,Moore1966,Neckel1984,Greisen2006,Bertaux2014} has also been included for comparison.\\
For HARPS DRS, the median SNR of the blue component is approx. $87$ and of the red component approx. $163$, for PyReduce $52$ and $145$, and for PyVesta $47$ and $132$, respectively. The lower SNR in the blue component is due to the significantly lower intensity of the solar spectrum in the blue spectral range. Both PyReduce and PyVesta have a drawback, particularly when the SNR is poor. It is unclear whether this is due to better optimization of the HARPS DRS or due to additional calibration data for the HARPS DRS that was not available through the ESO Archive.  

The median wavelength calibration error for the HARPS standard DRS is $9.86$ m/s across all orders; for PyVesta, it is $25.97$ m/s for the blue component and $27.81$ m/s for the red component. These errors represent the RMS  between the ThAr calibration line wavelengths and the fitted wavelength solution. Since PyVesta is not optimized for the ThAr line positions of HARPS but instead uses the more general wavelength calibration, its wavelength accuracy is slightly lower than that of the HARPS standard DRS. For our application, however, the accuracy is still more than sufficient. Wavelength calibration using PyReduce failed, so no wavelength comparison can be made between PyReduce and PyVesta. A comparison of the wavelength solutions from HARPS DRS and PyVesta is given in Fig. \ref{fig:Comparison_wavesolutions}. In general, a trend toward a positive shift can be observed in the picture. The median shift is approx. $8.19$ m/s, but it is significantly smaller than the mean error in the wavelength solution of about $26.9$ m/s (marked in red). It is also evident for the individual order that the median value of the residuals is not zero. The reason is that the fit for the wavelength solution is a 2D polynomial fit. This minimizes the residuals of all orders collectively, rather than those of a single order. As a result, individual orders may have a small offset.

\subsection{Discussion} \label{subsec:Discussion}
PyVesta offers several unique features compared to other data reduction software packages. IRAF/PyRAF and MIDAS are very old and quite difficult for new users to learn, so they are practically no longer used. PyReduce and PypeIt are more modern and are successfully used with various instruments. 

PyVesta now opens the door to easy use of low- to medium-resolution echelle spectrographs. Unlike the other packages, PyVesta is not designed to provide the most accurate analysis possible of high-resolution spectrographs such as HARPS and FEROS. Instead, PyVesta is intended to take advantage of the fact that off-the-shelf spectrographs serve as a good complement to high-resolution spectrographs by enabling the acquisition of time-consuming follow-up observations. Consistent data reduction using a single extraction software package is particularly advantageous when analyzing data from multiple instruments. In most cases, the off-the-shelf spectrograph will be the limiting factor, so there is no need to refine the extraction process for high-resolution spectrographs.

Instead, PyVesta places particular emphasis on the analysis of that off-the-shelf spectrographs. These spectrographs are generally less well characterized and have more optical aberrations than high-resolution spectrographs. Therefore, solutions have been developed to improve the extraction of spectra from these spectrographs and simplify wavelength calibration. However, these improvements also benefit high-resolution spectrographs, since even for those, not all necessary calibration data is always publicly available. 

Since PyVesta places a special focus on these spectrographs, their use at universities and even schools is greatly simplified, as an easily customizable data reduction software package is available now. This makes it easier to use these instruments for educational purposes without students having to learn how to use complicated software packages.

\subsection{Conclusion} \label{subsec:Conclusion}
Combining data from multiple echelle spectrographs offers significant advantages in the analysis of time-dependent spectra, such as the study of spectroscopic binary stars. To ensure consistent data analysis, it is important to process all data using data reduction software that is as universal as possible. CERES is an excellent choice for this purpose, but it is no longer up to date. PyVesta offers a new and easily customizable alternative with updated and refined methods, written in Python 3. 

PyVesta has currently been tested with the HARPS, FEROS, and Shelyak eShel II spectrographs. Additional spectrographs can easily be added. Instructions can be found on the PyVesta website. Additional code examples and test data are also available there.

\section*{Acknowledgments}
The authors thank MARA, the MArburg University Research Academy, for financial support. This paper uses observations made with ESO Telescopes at the La Silla Paranal Observatory under programme ID 088.C-0323(A). PyVesta makes use of numpy \citep{Harris2020} and scipy \citep{Virtanen2020}.

\bibliography{PyVesta}{}
\end{document}